\documentclass[11pt]{article}
\pdfoutput=1
\usepackage{jheppub}
\usepackage[T1]{fontenc}
\usepackage{physics}
\usepackage{bbm}
\usepackage{empheq}
\usepackage{tikz}
\usepackage{cancel}
\usepackage{xcolor,cancel}
\usepackage{hhline}

\newcommand\nn{\nonumber}
\newcommand\fft[2]{\frac{#1}{#2}}

\newcommand\mQ{\mathcal{Q}}

\newcommand\mR{\mathcal{R}}

\newcommand\mI{\mathcal{I}}
\newcommand\tz{\tilde{z}}
\newcommand\tu{\tilde{u}}

\newcommand\mn{\mathfrak{n}}
\newcommand\tmn{\tilde{\mathfrak{n}}}
\newcommand\mm{\mathfrak{m}}
\newcommand\tmm{\tilde{\mathfrak{m}}}
\newcommand\mft{\mathfrak{t}}

\newcommand\ts{\tilde{s}}
\newcommand\bs{\bar{s}}
\newcommand\bts{\bar{\tilde{s}}}
\newcommand\tU{\tilde{U}}
\newcommand\bU{\bar{U}}
\newcommand\btU{\bar{\tilde{U}}}
\newcommand\mW{\mathcal{W}}
\newcommand\mF{\mathcal{F}}

\newcommand\mV{\mathcal{V}}
\newcommand\tn{\tilde{n}}
\newcommand\tdh{\tilde{h}}
\newcommand\tx{\tilde{x}}
\newcommand\tq{\tilde{q}}
\newcommand\tDelta{\tilde{\Delta}}
\newcommand\ty{\tilde{y}}

\def\ri{{\rm i}}

\makeatletter
\newcommand*{\rom}[1]{\expandafter\@slowromancap\romannumeral #1@}
\makeatother

\preprint{KIAS-P22068}
\title{Large $N$ Superconformal Indices for 3d Holographic SCFTs}

\author[a]{Nikolay Bobev,}
\author[b]{Sunjin Choi,}
\author[a]{Junho Hong,}
\author[a]{and Valentin Reys}

\affiliation[a]{Institute for Theoretical Physics, KU Leuven\,,\\ Celestijnenlaan 200D, B-3001 Leuven, Belgium}
\affiliation[b]{School of Physics, Korea Institute for Advanced Study\,,\\ 85 Hoegiro, Dongdaemun-gu, Seoul 02455, Republic of Korea}

\emailAdd{nikolay.bobev@kuleuven.be}
\emailAdd{sunjinchoi@kias.re.kr}
\emailAdd{junho.hong@kuleuven.be}
\emailAdd{valentin.reys@kuleuven.be}
 
\abstract{We study a limit of the superconformal index of the ABJM theory on $S^1\times S^2$ in which the size of the circle is much smaller than the radius of the two-sphere. We derive closed form expressions for the two leading terms in this Cardy-like limit which are valid to all orders in the $1/N$ expansion. These results are facilitated by a judicious rewriting of the superconformal index which establishes a connection with the Bethe Ansatz Equations that control the topologically twisted index. Using the same technique we extend these results to the superconformal index of another holographic theory: 3d $\mathcal{N}=4$ SYM coupled to one adjoint and $N_f$ fundamental hypermultiplets. We discuss the implications of our results for holography and the physics of charged rotating black holes in AdS$_4$. }

\begin{document}

\maketitle


\section{Introduction}\label{sec:intro}

Understanding the properties of the spectrum of local BPS operators in superconformal field theories (SCFTs) offers a window into their strong coupling dynamics and is a subject with illustrious history. Supersymmetric indices \`a la Witten provide a convenient tool to encode such BPS spectra and are therefore objects of central interest in SCFTs. Such supersymmetric indices acquire new importance in the context of AdS/CFT and black hole physics where they can be used with great efficacy to do precision holography or to account for the microscopic entropy of black holes. Supersymmetric localization~\cite{Pestun:2007rz,Pestun:2016zxk} has been extensively used for exact calculations of path integrals in strongly interacting quantum field theories and as we discuss in this work can be brought to bear yet again for the calculation of supersymmetric indices in large $N$ holographic SCFTs.

The superconformal index (SCI) of the 4d $\mathcal N=4$ Super-Yang-Mills (SYM) theory on $S^1\times S^3$ defined in \cite{Kinney:2005ej,Romelsberger:2005eg} has received particular interest in this context. This is largely due to the recent observation that the SCI of the $\mathcal N=4$ SYM theory can account for the entropy of the dual supersymmetric Kerr-Newman (KN) AdS$_5$ black holes found in \cite{Gutowski:2004ez,Gutowski:2004yv,Chong:2005hr,Chong:2005da,Kunduri:2006ek}. The key to this microscopic understanding of the black hole entropy is to allow for complex chemical potentials associated with the global charges of the $\mathcal N=4$ SYM theory when evaluating the SCI~\cite{Hosseini:2017mds,Choi:2018hmj,Cabo-Bizet:2018ehj,Benini:2018ywd}. This can be interpreted as choosing the ``second sheet'' for the complex valued chemical potentials~\cite{Cassani:2021fyv}. These developments have sparked a flurry of activity centered at understanding the SCI of $\mathcal N=4$ SYM theory and its generalizations to other $\mathcal N=1$ holographic SCFTs that provide the microscopic origin of the entropy of dual AdS$_5$ black holes. A remarkable feature of the SCI in 4d $\mathcal N=1$ holographic SCFTs is that it allows for an all order expansion in a Cardy-like limit in which the size of the $S^1$ is taken to be much smaller than the radius of the $S^3$, see \cite{GonzalezLezcano:2020yeb,ArabiArdehali:2021nsx,Cassani:2021fyv} as well as \cite{Honda:2019cio,ArabiArdehali:2019tdm,Kim:2019yrz,Cabo-Bizet:2019osg,Amariti:2019mgp,ArabiArdehali:2019orz,Amariti:2020jyx,Ardehali:2021irq} for further discussion on the 4d SCI in this Cardy-like limit. In the large $N$ limit, on the other hand, the exact evaluation of the 4d SCI becomes much more involved due to multiple competing saddles; see \cite{GonzalezLezcano:2019nca,Lanir:2019abx,Cabo-Bizet:2019eaf,Cabo-Bizet:2020nkr,Benini:2020gjh,Copetti:2020dil,Goldstein:2020yvj,Cabo-Bizet:2020ewf,Choi:2021lbk,Gaiotto:2021xce,Choi:2021rxi,Choi:2022ovw,Bobev:2022bjm,Cassani:2022lrk} for a selection of recent results on various aspects of the SCI in this large $N$ limit. 

Our goal in this work is to study the large $N$ limit of the SCI on $S^1\times S^2$ defined and studied in \cite{Bhattacharya:2008zy,Bhattacharya:2008bja,Kim:2009wb,Imamura:2011su,Krattenthaler:2011da,Kapustin:2011jm} for 3d holographic SCFTs. The calculation of the 3d SCI in the large $N$ limit has been previously studied in a particular regime of fugacities but the result has been restricted to the leading order in the large $N$ limit taken after the Cardy-like limit \cite{Choi:2019zpz,Choi:2019dfu,Nian:2019pxj}.\footnote{See also \cite{GonzalezLezcano:2022hcf} for some very recent related work.} In the discussion below we focus on the Cardy-like limit, which ultimately amounts to taking the limit of vanishing $S^2$ angular momentum fugacity $\omega$, and improve on this analysis in two significant ways. Focusing on the ${\rm U}(N)_k\times {\rm U}(N)_{-k}$ ABJM theory for concreteness \cite{Aharony:2008ug}, we first rewrite the SCI in a suitable manner that simplifies the analysis of the leading $\omega^{-1}$ and $\omega^0$ terms in the Cardy-like limit.  We then use a saddle point approximation to show that the $\omega^{-1}$ term in the small $\omega$ expansion of the SCI is determined by the so-called Bethe potential used in the calculation of another related supersymmetric observable, the Topologically Twisted Index (TTI) \cite{Benini:2015noa,Benini:2016hjo,Closset:2016arn}. Moreover, we find that the  $\omega^0$ term in the Cardy-like limit simply evaluates to the TTI itself. Such a judicious rewriting of the SCI was studied in \cite{Choi:2019zpz,Choi:2019dfu} for general 3d $\mathcal{N}=2$ gauge theories where it was also shown that it can be used to compute the Cardy-like limit of the SCI for general values of $N$. These results are significant since the TTI has recently been computed to all orders in the $1/N$ expansion for various holographic SCFTs, including the ABJM theory~\cite{Bobev:2022jte,Bobev:2022eus,Bobev:2023lkx}. We can thus leverage these recent TTI results to obtain a closed form expression for the $\omega^{-1}$ and $\omega^0$ terms in the Cardy-like limit of the SCI for the ABJM theory to all orders in the $1/N$ expansion at fixed $k$. We believe that our results can be extended to many holographic 3d $\mathcal{N}=2$ SCFTs. To illustrate this we use the same techniques to compute explicitly the $\omega^{-1}$ and $\omega^0$ terms in the Cardy-like limit of the SCI for 3d ${\rm U}(N)$ $\mathcal{N}=4$ SYM coupled to one adjoint and $N_f$ fundamental hypermultiplets\footnote{This model is also known as the ADHM quiver theory and we will often adopt this monicker.} to all orders in the $1/N$ expansion at fixed $N_f$. The $S^3$ partition function and the TTI of these two holographic SCFTs can also be computed in a closed form to all orders in the large $N$ expansion, see \cite{Fuji:2011km,Marino:2011eh,Mezei:2013gqa,Grassi:2014vwa} and \cite{Bobev:2022jte,Bobev:2023lkx}, respectively. Our findings thus add to the menagerie of exact large $N$ results in these two holographic models.

The 3d SCI for holographic SCFTs arising from M2-branes can be used to understand the leading $N^\fft32$ term in the large $N$ limit of the entropy of the dual supersymmetric Kerr-Newman black holes in AdS$_4$ constructed in \cite{Kostelecky:1995ei,Caldarelli:1998hg,Cvetic:2005zi,Chow:2013gba,Hristov:2019mqp}, see for instance \cite{Choi:2018fdc,Cassani:2019mms,Hristov:2019mqp}. Our explicit results for the SCI go well beyond this leading order in the large $N$ limit and should thus have interesting holographic implications. In particular, we find that the $N^\fft12$ term in the large $N$ expansion of our SCI results agrees with the holographically dual calculation performed in \cite{Bobev:2020egg,Bobev:2021oku} using higher-derivative supergravity. Moreover, we find that the $\log N$ term in the large $N$ expansion of the SCI agrees with the dual 1-loop calculation in supergravity discussed in \cite{Hristov:2021zai}. We also show that our results are compatible with a recent conjecture \cite{Hristov:2022lcw}, motivated by 4d gauged supergravity and holography, for a closed form expression for the SCI of the ABJM theory to all orders in the $1/N$ expansion. Finally, we discuss how the subleading $1/N$ corrections to the 3d SCI that we have obtained from the field theory side can improve on the previous microstate counting of the entropy of the holographic dual supersymmetric AdS$_4$ black hole.

The rest of this paper is organized as follows. In Section~\ref{sec:general} we briefly review the SCI of $\mathcal N=2$ SCFTs and the corresponding matrix model obtained via supersymmetric localization. In Sections~\ref{sec:ABJM} and \ref{sec:ADHM} we investigate the SCI of the ABJM theory and the ADHM theory respectively, and provide the all order perturbative $1/N$ expansion for the first two leading terms in the Cardy-like expansion. In Section~\ref{sec:holo} we discuss the holographic implications of our results and their bearing on the thermodynamics of supersymmetric Kerr-Newman AdS$_4$ black holes. We conclude in Section~\ref{sec:discussion} with a short discussion of some open questions. The four appendices contain a collection of technical results on special functions and the saddle point approximation along with a summary of the numerical analysis used to derive our main results.

\section{Superconformal index of generic $\mathcal N=2$ SCFTs}\label{sec:general}

The $S^1\times S^2$ superconformal index (SCI) was defined in \cite{Bhattacharya:2008zy,Bhattacharya:2008bja,Kim:2009wb} and then considered for generic $\mathcal N=2$ SCFTs in \cite{Imamura:2011su,Krattenthaler:2011da,Kapustin:2011jm}. It can be viewed as the following trace over the Hilbert space of the theory in radial quantization
\begin{equation}
	\mI(q,\xi_a)=\Tr\relax\bigg[(-1)^Fe^{-\beta_1\{\mQ,\mQ^\dagger\}}q^{\Delta+j_3}\prod_{a}\xi_a^{F_a}\bigg]\,,\label{SCI:tr:1}
\end{equation}
where $F$ is the fermion number and the supercharge $\mQ$ satisfies the anticommutation relation
\begin{equation}
	\{\mQ,\mQ^\dagger\}=\Delta-R-j_3\,.
\end{equation}
The charges $\Delta+j_3$ and $F_a$ in the trace formula (\ref{SCI:tr:1}) commute with $\mQ$ and $\mQ^\dagger$. Here $\Delta$ is the energy in radial quantization, $R$ denotes the superconformal $R$-charge, $j_3$ is the third component of the angular momentum on $S^2$, and $F_a$ are charges associated with flavor symmetries. 

We use the conventions of \cite{Imamura:2011su} where the ratio of the circumference of $S^1$ and the radius of $S^2$ is denoted by $\beta=\beta_1+\beta_2$. The SCI (\ref{SCI:tr:1}) is independent of the parameter $\beta_1$ based on the usual argument for the Witten index and receives contributions only from the BPS states annihilated by $\mQ$ and $\mQ^\dagger$ or equivalently the BPS states saturating the bound $\Delta-R-j_3\geq0$. As a result one can rewrite the SCI (\ref{SCI:tr:1}) as
\begin{equation}
	\mI(q,\xi_a)=\Tr_\text{BPS}\bigg[(-1)^Fq^{R+2j_3}\prod_{a}\xi_a^{F_a}\bigg]\,,\label{SCI:tr:2}
\end{equation}
where $\Tr_\text{BPS}[\cdots]$ means that the trace is taken over the BPS states annihilated by $\mQ$ and $\mQ^\dagger$ only. We will call the chemical potential for the $q$ fugacity  $\omega$ and will use the relation $q=e^{\ri\pi\omega}$. The relation between the geometric parameter $\beta_2$ used in \cite{Imamura:2011su} and $\omega$ is then $\pi \omega = \ri \beta_2$. Since the SCI is independent of $\beta_1$ we can take it to vanish and then use $\beta_2$ as the real parameter that determines the length of the $S^1$ (in units of the $S^2$ radius). Therefore the Cardy-like limit of small $S^1$ corresponds to taking $\omega \to i0^+$. This limit will play a central role in our discussion below.

The SCI (\ref{SCI:tr:2}) can also be written explicitly in terms of a matrix model for generic $\mathcal N=2$ SCFTs \cite{Imamura:2011su,Krattenthaler:2011da,Kapustin:2011jm}. In this paper we focus on the case where a given $\mathcal N=2$ SCFT has the gauge group $G=\otimes_{r=1}^p\text{U}(N)_{k_r}$ with Chern-Simons (CS) levels $k_r$'s and $\mathcal N=2$ chiral multiplets collectively represented by $\Phi$ with superconformal $R$-charge $R(\Phi)$ and flavor charges $f_a(\Phi)$. The matrix model for the SCI then reads\footnote{We mainly followed the conventions of Appendix A in \cite{Aharony:2013dha} involving the extra phases in the 2nd line of (\ref{SCI}) compared to the previous localization formula in the literature \cite{Imamura:2011su,Krattenthaler:2011da,Kapustin:2011jm}, which comes from the canonical prescription that the fermionic number operator $(-1)^F$ in the trace formula (\ref{SCI:tr:1}) is understood as $(-1)^F=(-1)^{2j_3}$. See also \cite{Dimofte:2011py,Aharony:2013kma} for related discussions.} 
\begin{align}\label{SCI}
	\mI(q,\xi_a)&=\fft{1}{(N!)^p}\sum_{\mm_1,\cdots,\mm_p\in\mathbb Z^N}\oint\bigg(\prod_{r=1}^p\prod_{i=1}^N\fft{dz_{r,i}}{2\pi \ri z_{r,i}}(z_{r,i})^{k_r\mm_{r,i}}\prod_a\xi_a^{k_{a,r,i}\mm_{r,i}}\bigg)\notag\\
	&\quad\times\prod_{r=1}^p\prod_{i=1}^N(-1)^{k_r\mm_{r,i}^2}\times\prod_\Phi\prod_\rho(-1)^{\fft12\rho(\mm)|\rho(\mm)|}\nn\\
	&\quad\times\prod_{r=1}^p\prod_{i\neq j}^Nq^{-\fft12|\mm_{r,i}-\mm_{r,j}|}\left(1-z_{r,i}z_{r,j}^{-1}q^{|\mm_{r,i}-\mm_{r,j}|}\right)\\
	&\quad\times\prod_{\Phi}\prod_{\rho}\left(q^{1-R(\Phi)}e^{-\ri\rho(h)}\prod_{a}\xi_a^{-f_a(\Phi)}\right)^{\fft12|\rho(\mm)|}\fft{(e^{-\ri\rho(h)}\prod_a\xi_a^{-f_a(\Phi)}q^{2-R(\Phi)+|\rho(\mm)|};q^2)_\infty}{(e^{\ri\rho(h)}\prod_a\xi_a^{f_a(\Phi)}q^{R(\Phi)+|\rho(\mm)|};q^2)_\infty}\,,\notag
\end{align}
where the contour integrals for the gauge zero modes $z_{r,i}=e^{\ri h_{r,i}}$ are over the unit circles and $\rho$ runs over the weights of the representation $\mR_\Phi$ of the $\mathcal N=2$ chiral multiplet $\Phi$ with respect to the gauge group $G=\otimes_{r=1}^p\text{U}(N)_{k_r}$. The $\mm_{r,i}$ denote quantized gauge magnetic fluxes. In Appendix \ref{App:special} we provide the definition of the $\infty$-Pochhammer symbol $(\cdot;\cdot)_\infty$ and present some of its properties.

It is worth mentioning that in the matrix model (\ref{SCI}) we have also included the contribution from mixed CS terms between gauge and flavor symmetries with general CS levels $k_{a,r,i}$ following Appendix A of \cite{Aharony:2013dha}. For the concrete examples in the following sections, however, we will turn on such mixed CS terms only for the topological symmetries with unit CS levels by coupling a background vector multiplet to the topological symmetry current $J_{T_r}=*\Tr[F_r]$ with gauge field strength $F_r$.

\section{ABJM superconformal index}\label{sec:ABJM}

In this section we investigate the SCI of the ABJM theory. In Section~\ref{sec:ABJM:mm} we briefly introduce the ABJM theory \cite{Aharony:2008ug} and then provide the matrix model for its SCI based on the general presentation in Section~\ref{sec:general}. In Section~\ref{sec:ABJM:Cardy} we evaluate the matrix model for the ABJM SCI using the saddle point approximation in the Cardy-like limit following \cite{Choi:2019zpz,Choi:2019dfu}. In Section~\ref{sec:ABJM:largeN} we provide the all order perturbative $1/N$ expansion for the first two leading terms in the Cardy-like expansion of the ABJM SCI using the exact ABJM TTI on $S^1\times S^2$ recently found in the literature \cite{Bobev:2022jte,Bobev:2022eus}. Finally, in Section~\ref{sec:ABJM:Airy} we discuss the relation between the all order results for the ABJM SCI and the Airy conjecture recently proposed in \cite{Hristov:2021qsw,Hristov:2022lcw}.

\subsection{Matrix model for the ABJM SCI}\label{sec:ABJM:mm}

The ABJM theory has U$(N)_k\times$U$(N)_{-k}$ gauge group with opposite CS levels $\pm k$, the corresponding two $\mathcal N=2$ vector multiplets, and two pairs of bi-fundamental and anti bi-fundamental $\mathcal N=2$ chiral multiplets $A_{1,2}$ \& $B_{1,2}$ with superconformal $R$-charge $\fft12$. We use $\Phi$ to collectively represent all the $\mathcal N=2$ chiral multiplets as $\Phi=\{A_1,A_2,B_1,B_2\}$.

In addition to the U$(1)_R$ superconformal R-symmetry, the global symmetry of the ABJM theory manifest in this $\mathcal{N}=2$ formulation is~SU(2)$_A\times$SU(2)$_B\times$U(1)$_T$. The flavor symmetries SU(2)$_A$ and SU(2)$_B$ act on $A_{1,2}$ and $B_{1,2}$ respectively, while U(1)$_T$ denotes the topological symmetry arising from the U$(1)$ factors in the gauge groups. We denote the Cartan generators of these global symmetries as $(A,B,T,R)$ respectively. The charges of the $\mathcal N=2$ chiral multiplets under these symmetries are presented in Table~\ref{ABJM:global}, see for example \cite{Benini:2015eyy} for more details.
\begin{table}
	\centering
	\begin{tabular}{|c||c|c|c|c|}
		\hline
		$\Phi$ & $R$ & $A$ & $B$ & $T$ \\\hhline{|=||=|=|=|=|}
		$A_1$ & $\fft12$ & $1$ & $0$ & $0$\\\hline
		$A_2$ & $\fft12$ & $-1$ & $0$ & $0$\\\hline
		$B_1$ & $\fft12$ & $0$ & $1$ & $0$\\\hline
		$B_2$ & $\fft12$ & $0$ & $-1$ & $0$\\\hline
	\end{tabular}
	\caption{Charge assignments of $\mathcal N=2$ chiral multiplets in the ABJM theory. The U$(1)_T$ symmetry acts non-trivially on monopole operators.}
	\label{ABJM:global}
\end{table}

The ABJM SCI can be obtained from the trace formula \eqref{SCI:tr:2} and reads
\begin{equation}
	\mI_\text{ABJM}(q,\xi_A,\xi_B,\xi_T)=\Tr_\text{BPS}\bigg[(-1)^Fq^{R+2j_3}(\xi_A)^{A}(\xi_B)^{B}(\xi_T)^T\bigg]\,,\label{ABJM:SCI:tr}
\end{equation}
where $(\xi_A,\xi_B,\xi_T)$ are fugacities associated with the flavor charges $(A,B,T)$ respectively. The corresponding matrix model for the ABJM SCI can then be obtained from \eqref{SCI} using Table~\ref{ABJM:global}
\begin{equation}
	\begin{split}
		&\mI_\text{ABJM}(q,\xi_A,\xi_B,\xi_T)\\
		&=\fft{1}{(N!)^2}\sum_{\mm,\tmm\in\mathbb Z^N}\oint\bigg(\prod_{i=1}^N\fft{dz_i}{2\pi \ri z_i}\fft{d\tz_i}{2\pi \ri \tz_i}z_i^{ k\mm_i}\tz_i^{-k\tmm_i}\xi_{T_1}^{\mm_i}\xi_{T_2}^{\tmm_i}\bigg)\\
		&\quad\times\prod_{i\neq j}^Nq^{-\fft12|\mm_i-\mm_j|}\left(1-z_iz_j^{-1}q^{|\mm_i-\mm_j|}\right)q^{-\fft12|\tmm_i-\tmm_j|}\left(1-\tz_i\tz_j^{-1}q^{|\tmm_i-\tmm_j|}\right)\\
		&\quad\times\prod_{i,j=1}^N\left(q^{\fft12}z_i^{-1}\tz_j\xi_A^{-1}\right)^{\fft12|\mm_i-\tmm_j|}\fft{(z_i^{-1}\tz_j\xi_A^{-1}q^{\fft32+|-\mm_i+\tmm_j|};q^2)_\infty}{(z_i\tz_j^{-1}\xi_Aq^{\fft12+|\mm_i-\tmm_j|};q^2)_\infty}\\
		&\quad\times\prod_{i,j=1}^N\left(q^{\fft12}z_i^{-1}\tz_j\xi_A\right)^{\fft12|\mm_i-\tmm_j|}\fft{(z_i^{-1}\tz_j\xi_Aq^{\fft32+|-\mm_i+\tmm_j|};q^2)_\infty}{(z_i\tz_j^{-1}\xi_A^{-1}q^{\fft12+|\mm_i-\tmm_j|};q^2)_\infty}\\
		&\quad\times\prod_{i,j=1}^N\left(q^{\fft12}z_i\tz_j^{-1}\xi_B^{-1}\right)^{\fft12|\mm_i-\tmm_j|}\fft{(z_i\tz_j^{-1}\xi_B^{-1}q^{\fft32+|-\mm_i+\tmm_j|};q^2)_\infty}{(z_i^{-1}\tz_j\xi_Bq^{\fft12+|\mm_i-\tmm_j|};q^2)_\infty}\\
		&\quad\times\prod_{i,j=1}^N\left(q^{\fft12}z_i\tz_j^{-1}\xi_B\right)^{\fft12|\mm_i-\tmm_j|}\fft{(z_i\tz_j^{-1}\xi_Bq^{\fft32+|-\mm_i+\tmm_j|};q^2)_\infty}{(z_i^{-1}\tz_j\xi_B^{-1}q^{\fft12+|\mm_i-\tmm_j|};q^2)_\infty}\,,
	\end{split}\label{ABJM:SCI:1}
\end{equation}
where we have turned on the mixed CS terms between the gauge symmetry and the topological symmetries only. Note that the ABJM SCI (\ref{ABJM:SCI:1}) does not depend on $\xi_{T_1}$ and $\xi_{T_2}$ separately but only on $\xi_T\equiv \xi_{T_1}\xi_{T_2}$ as expected from the fact that ABJM theory has only one non-trivial U(1)$_T$ topological symmetry \cite{Aharony:2008ug}. This can be seen explicitly in the matrix model \eqref{ABJM:SCI:1} by shifting the integration variables as $(z_i,\tz_i)\to(z_i\xi_{T_1}^{-1/k},\tz_i\xi_{T_2}^{1/k})$. The phases in the 2nd line of (\ref{SCI}) disappear in the ABJM SCI (\ref{ABJM:SCI:1}) since the phases from bi-fundamental chiral multiplets cancel each other and the phase from CS terms also reduces to unity for an integer CS level $k$.\footnote{To be rigorous, the phase from the CS terms $(-1)^{k(\mm_i^2-\tmm_i^2)}$ may not always be trivial if the CS level $k$ is odd. However, we do not expect that this subtle phase factor involves important physics that is sensitive to the parity of the CS level. We leave a thorough investigation of this issue for the future, and refer the reader to \cite{Dimofte:2011py,Hwang:2012jh} for relevant discussion on non-trivial phase factors in the localization of SCI.}

To study the Cardy-like limit of the ABJM SCI it is useful to rewrite the matrix model (\ref{ABJM:SCI:1}) as \cite{Choi:2019zpz,Choi:2019dfu}
\begin{equation}
	\begin{split}
		&\mI_\text{ABJM}(\omega,\Delta,\mn)\\
		&=\fft{1}{(N!)^2}\sum_{\mm,\tmm\in\mathbb Z^N}\oint_{|s_i|=q^{\mm_i},|\ts_i|=q^{\tmm_i}}\bigg(\prod_{i=1}^N\fft{ds_i}{2\pi \ri s_i}\fft{d\ts_i}{2\pi \ri\ts_i}e^{\fft{\ri k}{4\pi\omega}\left(U_i^2-\tU_i^2\right)-\fft{\ri k}{4\pi\omega}\left(\bU_i^2-\btU_i^2\right)}\bigg)\\
		&\quad\times\prod_{i\neq j}^N\Big[(1-s_i^{-1}s_j)(1-\ts_i^{-1}\ts_j)(1-\bs_i^{-1}\bs_j)(1-\bts_i^{-1}\bts_j)\Big]^\fft12\\
		&\quad\times\prod_{i=1}^N\Bigg[(s_i\bs_i)^{\fft{N}{2}-i}(\ts_i\bts_i)^{\fft{N}{2}-i+1}\prod_{a=1}^2\fft{(s_i^{-1}\ts_iy_a^{-1}q^{2-\mn_a};q^2)_\infty}{(\bs_i^{-1}\bts_iy_aq^{\mn_a};q^2)_\infty}\times\prod_{a=3}^4\fft{(\bs_i^{-1}\bts_iy_a^{-1}q^{2-\mn_a};q^2)_\infty}{(s_i^{-1}\ts_iy_aq^{\mn_a};q^2)_\infty}\Bigg]\\
		&\quad\times\prod_{i>j}^N\Bigg[\prod_{a=1}^2\fft{(s_i^{-1}\ts_jy_a^{-1}q^{2-\mn_a};q^2)_\infty}{(\bs_i^{-1}\bts_jy_aq^{\mn_a};q^2)_\infty}\times\prod_{a=3}^4\fft{(\bs_i^{-1}\bts_jy_a^{-1}q^{2-\mn_a};q^2)_\infty}{(s_i^{-1}\ts_jy_aq^{\mn_a};q^2)_\infty}\\
		&\kern4em~\times\prod_{a=1}^2\fft{(\bs_j\bts_i^{-1}y_a^{-1}q^{2-\mn_a};q^2)_\infty}{(s_j\ts_i^{-1}y_aq^{\mn_a};q^2)_\infty}\times\prod_{a=3}^4\fft{(s_j\ts_i^{-1}y_a^{-1}q^{2-\mn_a};q^2)_\infty}{(\bs_j^{-1}\bts_iy_aq^{\mn_a};q^2)_\infty}\Bigg]\,,
	\end{split}\label{ABJM:SCI:3}
\end{equation}
in terms of new integration variables (recall that $z_i=e^{\ri h_i},\,\tz_i=e^{\ri\tdh_i},\,q=e^{\ri\pi\omega}$) 
\begin{equation}
	\begin{alignedat}{2}
		s_i&=e^{\ri U_i}=z_iq^{\mm_i}=e^{\ri (h_i+\pi\omega\mm_i)}\,,&\qquad \ts_i&=e^{\ri\tU_i}=\tz_iq^{\tmm_i}=e^{\ri(\tdh_i+\pi\omega\tmm_i)}\,,
	\end{alignedat}\label{ABJM:s}
\end{equation}
and the parameters $y_a=e^{\ri\pi\Delta_a}\,(\Delta_a\in\mathbb R)$ and $\mn_a\in\mathbb R$ defined as
\begin{equation}
	y_aq^{\mn_a-\fft12}=(\xi_T^{-1/k}\xi_A,\xi_T^{-1/k}\xi_A^{-1},\xi_T^{1/k}\xi_B,\xi_T^{1/k}\xi_B^{-1})\qquad(a\in\{1,2,3,4\})\,,\label{ABJM:yn}
\end{equation}
under the constraints
\begin{equation}
	\sum_{a=1}^4\mn_a=2\,,\qquad\prod_{a=1}^4y_a=1\,.\label{ABJM:constraint}
\end{equation}
Note that $(y_a,\mn_a)$ are uniquely determined for given three complex parameters $(\xi_A,\xi_B,\xi_T)$ through the map (\ref{ABJM:yn}) under the constraints (\ref{ABJM:constraint}). Accordingly in (\ref{ABJM:SCI:3}) we have replaced the argument of the ABJM SCI as
\begin{equation}
	\mI_\text{ABJM}(q,\xi_A,\xi_B,\xi_T)\quad\to\quad\mI_\text{ABJM}(\omega,\Delta,\mn)\,.
\end{equation}
where $(\Delta,\mn)$ collectively represent $(\Delta_a,\mn_a)$ under the constraints (\ref{ABJM:constraint}). To obtain the expression (\ref{ABJM:SCI:3}) we have also assumed
\begin{equation}
	q\in\mathbb R_{>0}\quad\Leftrightarrow\quad \ri\omega\in\mathbb R\,.
\end{equation}
In Appendix \ref{App:ABJM:Cardy}, we provide some more details on the derivation of (\ref{ABJM:SCI:3}).

\subsection{Cardy-like limit of the ABJM SCI}\label{sec:ABJM:Cardy}

Taking the Cardy-like limit
\begin{equation}
	|q|\to1^-\quad\Leftrightarrow\quad \omega\to i0^+\,,\label{Cardy}
\end{equation}
one can expand the $\infty$-Pochhammer symbol using the asymptotic expansion (\ref{poch:asymp}) and apply the Euler-Maclaurin formula to replace the sum over gauge magnetic fluxes with the corresponding integrals in the matrix model (\ref{ABJM:SCI:3}), see \cite{Choi:2019zpz,Choi:2019dfu} and Appendix \ref{App:ABJM:Cardy} for details. The result reads
\begin{equation}
	\begin{split}
		&\mI_\text{ABJM}(\omega,\Delta,\mn)\\
		&=\fft{1}{(N!)^2}\int_{\mathbb C^{2N}}\bigg(\prod_{i=1}^N\fft{dU_id\bU_i}{-4\pi^2\ri\omega}\fft{d\tU_id\btU_i}{-4\pi^2\ri\omega}\bigg)\,e^{\fft{1}{\pi\omega}\Im\mW^{(0)}[U,\tU;\Delta]+2\Re\mW^{(1)}[U,\tU;\Delta,\mn]+\mathcal O(\omega)}\,,
	\end{split}\label{ABJM:SCI:Cardy:1}
\end{equation}
where the first two leading terms of the effective action in the Cardy-like limit are given by
\begin{subequations}
	\begin{align}
		\mathcal W^{(0)}[U,\tU;\Delta]&=-\fft{k}{2}\sum_{i=1}^N\left(U_i^2-\tU_i^2\right)+\sum_{i=1}^N\Bigg[\sum_{a=1}^2\text{Li}_2(s_i^{-1}\ts_iy_a^{-1})-\sum_{a=3}^4\text{Li}_2(s_i^{-1}\ts_iy_a)\Bigg]\nn\\
		&\quad+\sum_{i>j}\Bigg[\sum_{a=1}^2\bigg(\text{Li}_2(s_i^{-1}\ts_jy_a^{-1})-\text{Li}_2(s_j\ts_i^{-1}y_a)\bigg)\nn\\
		&\kern4em~~-\sum_{a=3}^4\bigg(\text{Li}_2(s_i^{-1}\ts_jy_a)-\text{Li}_2(s_j\ts_i^{-1}y_a^{-1})\bigg)\Bigg]\,,\label{ABJM:W:0}\\
		\mathcal W^{(1)}[U,\tU;\Delta,\mn]&=\fft12\sum_{i=1}^N\Bigg[\sum_{a=1}^2(1-\mn_a)\text{Li}_1(s_i^{-1}\ts_iy_a^{-1})+\sum_{a=3}^4(1-\mn_a)\text{Li}_1(s_i^{-1}\ts_iy_a)\Bigg]\nn\\
		&\quad+\fft12\sum_{i>j}\Bigg[\sum_{a=1}^2(1-\mn_a)\bigg(\text{Li}_1(s_i^{-1}\ts_jy_a^{-1})+\text{Li}_1(s_j\ts_i^{-1}y_a)\bigg)\nn\\
		&\kern5em+\sum_{a=3}^4(1-\mn_a)\bigg(\text{Li}_1(s_i^{-1}\ts_jy_a)+\text{Li}_1(s_j\ts_i^{-1}y_a^{-1})\bigg)\Bigg]\nn\\
		&\quad+\fft12\sum_{i\neq j}^N\bigg[\log(1-s_i^{-1}s_j)+\log(1-\ts_i^{-1}\ts_j)\bigg]\nn\\
		&\quad+\fft12\sum_{i=1}^N\bigg[(N-2i)\log s_i+(N-2i+2)\log\ts_i\bigg]\nn\\
		&\quad-\fft{N}{4}\Bigg[\sum_{a=1}^2(1-\mn_a)\log y_a-\sum_{a=3}^4(1-\mn_a)\log y_a\Bigg]\,.\label{ABJM:W:1}
	\end{align}\label{ABJM:W}%
\end{subequations}
Note that in (\ref{ABJM:SCI:Cardy:1}) we have used the complex conjugation relations
\begin{equation}
\begin{split}
	\overline{\mW^{(0)}[U,\tU;\Delta]}&=\mW^{(0)}[-\bU,-\btU;-\Delta]\,,\\
	\overline{\mW^{(1)}[U,\tU;\Delta,\mn]}&=\mW^{(1)}[-\bU,-\btU;-\Delta,\mn]\,,
\end{split}
\end{equation}
which are derived using that $(\Delta,\mn)$ are real by construction, see (\ref{ABJM:yn}).

To apply the saddle point approximation to the integral (\ref{ABJM:SCI:Cardy:1}), first one should solve the saddle point equations in the Cardy-like limit, namely
\begin{equation}
	0=\fft{\partial\mathcal W^{(0)}[U,\tU;\Delta]}{\partial U_i}=\fft{\partial\mathcal W^{(0)}[U,\tU;\Delta]}{\partial\tU_i}\,.\label{ABJM:Cardy:saddle}
\end{equation}
Let us denote the solution to the leading order saddle point equation (\ref{ABJM:Cardy:saddle}) by $\{U_{\star i},\tU_{\star i}\}$ with the apparent $(N!)^2$ degeneracy from permutations. The saddle point approximation to the integral (\ref{ABJM:SCI:Cardy:1}) can then be derived following Appendix \ref{App:saddle} and the result reads
\begin{equation}
\begin{split}
	&\log\mI_\text{ABJM}(\omega,\Delta,\mn)\\
	&=\fft{1}{\pi\omega}\Im\mW^{(0)}[U_\star,\tU_\star;\Delta]\\
	&\quad+\log\Bigg|\fft{\prod_{a=1}^4y_a^{-\fft{N^2}{2}\mn_a}}{\det\mathbb B[U_\star,\tU_\star;\Delta]}\fft{\prod_{i=1}^Ns_{\star i}^{N}\ts_{\star i}^{N}\prod_{i\neq j}\left(1-s_{\star i}^{-1}s_{\star j}\right)\left(1-\ts_{\star i}^{-1}\ts_{\star j}\right)}{\prod_{i,j=1}^N\prod_{a=1}^2(\ts_{\star i}-y_as_{\star j})^{1-\mn_a}\prod_{a=3}^4(s_{\star i}-y_a\ts_{\star j})^{1-\mn_a}}\Bigg|\\
	&\quad+\mathcal O(\omega)\,,
\end{split}\label{ABJM:SCI:Cardy:2}
\end{equation}
where the $2N\times 2N$ square matrix $\mathbb B[U,\tU;\Delta]$ is defined by its components as
\begin{equation}
	\Big[\mathbb B[U,\tU;\Delta]\Big]_{I,J}\equiv\fft{\partial\mW^{(0)}[U,\tU;\Delta]}{\partial U_I\partial U_J}\quad~(U_I=\{U_1,\cdots,U_N,\tU_1,\cdots,\tU_N\})\,.\label{ABJM:B:SCI}
\end{equation}
See Appendix \ref{App:ABJM:Cardy} for additional details.

\subsection{All order $1/N$ expansion for the ABJM SCI in the Cardy-like limit}\label{sec:ABJM:largeN}
To evaluate the Cardy-like expansion of the ABJM SCI (\ref{ABJM:SCI:Cardy:2}) explicitly in the large $N$ limit, we will use the known all order results for the $S^1\times S^2$ TTI of the ABJM theory \cite{Bobev:2022jte,Bobev:2022eus}. As a first step, we relabel the gauge holonomies and the leading order effective action in (\ref{ABJM:SCI:Cardy:2}) as
\begin{equation}
\begin{split}
	U_i\quad&\to\quad-\tu_i-\fft{(1-(-1)^N)\pi}{2k}\,,\\
	\tU_i\quad&\to\quad-u_i-\fft{(1-(-1)^N)\pi}{2k}\,,\\
	\mW^{(0)}[U,\tU;\Delta]\quad&\to\quad-\mV_\text{TTI}[u,\tu;\Delta]\,.\label{ABJM:SCI:to:TTI}
\end{split}
\end{equation}
Then by using the inversion formula of the polylogarithm (\ref{polylog:inversion}) and assuming
\begin{equation}
	0<\Re[\tu_i-u_j+\pi\Delta_{3,4}]<2\pi\,,\quad -2\pi<\Re[\tu_i-u_j-\pi\Delta_{1,2}]<0\qquad(i<j)\,,\label{ABJM:Del:range}
\end{equation}
one can rewrite $\mV_\text{TTI}[u,\tu;\Delta]$ as
\begin{equation}
\begin{split}
	&\mV_\text{TTI}[u,\tu;\Delta]\\
	&=\sum_{i=1}^N\Bigg[\fft{k}{2}\left(\tu_i^2-u_i^2\right)-\pi\left(2\tn_i-\fft{1-(-1)^N}{2}\right)\tu_i+\pi\left(2n_i-\fft{1-(-1)^N}{2}\right)u_i\Bigg]\\
	&\quad+\sum_{i,j=1}^N\Bigg[\sum_{a=3,4}\text{Li}_2(e^{\ri(\tu_j-u_i+\pi\Delta_a)})-\sum_{a=1,2}\text{Li}_2(e^{\ri(\tu_j-u_i-\pi\Delta_a)})\Bigg]\\
	&\quad-\fft{N(N-1)\pi^2}{4}\Bigg[\sum_{a=3}^4(1-\Delta_a)^2-\sum_{a=1}^2(1-\Delta_a)^2\Bigg]\,,
\end{split}\label{ABJM:V}
\end{equation}
where we have introduced a set of integers $(n_i,\tn_i)=(1-i,i-n)$ in the last equation. It is remarkable that (\ref{ABJM:V}) is exactly the same as the Bethe potential for the $S^1\times S^2$ ABJM TTI~\cite{Bobev:2022eus} up to $u,\tu$-independent terms. This implies that the leading order saddle point equations for the matrix model of the ABJM SCI in the Cardy-like limit, (\ref{ABJM:Cardy:saddle}), are equivalent to the Bethe Ansatz Equations (BAE) obtained by taking partial derivatives of the Bethe potential with respect to $(u_i,\tu_i)$,
\begin{equation}
	0=\fft{\partial\mV_\text{TTI}[u,\tu;\Delta]}{\partial u_i}=\fft{\partial\mV_\text{TTI}[u,\tu;\Delta]}{\partial\tu_i}\,.\label{ABJM:TTI:BAE}
\end{equation}
Based on the equivalence between the saddle point equation (\ref{ABJM:Cardy:saddle}) and the BAE (\ref{ABJM:TTI:BAE}) under the map (\ref{ABJM:SCI:to:TTI}), one can rewrite the Cardy expansion of the ABJM SCI (\ref{ABJM:SCI:Cardy:2}) as
\begin{equation}
\begin{split}
	&\log\mI_\text{ABJM}(\omega,\Delta,\mn)\\
	&=-\fft{1}{\pi\omega}\Im\mV_\text{TTI}[u_\star,\tu_\star;\Delta]\\
	&\quad+\log\Bigg|\fft{\prod_{a=1}^4y_a^{-\fft{N^2}{2}\mn_a}}{\det\mathbb B_\text{TTI}[u_\star,\tu_\star;\Delta]}\fft{\prod_{i=1}^Nx_{\star i}^{N}\tx_{\star i}^{N}\prod_{i\neq j}\left(1-x_{\star i}^{-1}x_{\star j}\right)\left(1-\tx_{\star i}^{-1}\tx_{\star j}\right)}{\prod_{i,j=1}^N\prod_{a=1}^2(\tx_{\star i}-y_ax_{\star j})^{1-\mn_a}\prod_{a=3}^4(x_{\star i}-y_a\tx_{\star j})^{1-\mn_a}}\Bigg|\\
	&\quad+\mathcal O(\omega)\,,
\end{split}\label{ABJM:SCI:Cardy:3}
\end{equation}
in terms of the solutions $\{x_{\star i}=e^{\ri u_{\star i}},\tx_{\star i}=e^{\ri\tu_{\star i}}\}$ to the BAE (\ref{ABJM:TTI:BAE}) and the $2N\times 2N$ matrix
\begin{equation}
	\Big[\mathbb B_\text{TTI}[u,\tu;\Delta]\Big]_{I,J}\equiv(-1)^{1-\Theta(I-N)}\fft{\partial\mV_\text{TTI}[u,\tu;\Delta]}{\partial u_I\partial u_J}\quad~(u_I=\{u_1,\cdots,u_N,\tu_1,\cdots,\tu_N\})\,,\label{ABJM:TTI:B}
\end{equation}
which is equivalent to the Jacobian matrix in the Bethe Ansatz formulation for the ABJM TTI \cite{Benini:2015eyy,Bobev:2022jte,Bobev:2022eus}. In (\ref{ABJM:TTI:B}) we have used the Heaviside step function $\Theta(x)$ on the real line that takes the value $\Theta(x)=1$ for $x>0$ and vanishes for $x\leq 0$.

Note that the first subleading term of order $\mathcal O(\omega^0)$ in (\ref{ABJM:SCI:Cardy:3}) is precisely the same as the logarithm of the $S^1\times S^2$ ABJM TTI, which has recently been evaluated to all orders in the perturbative $1/N$ expansion by using precision numerical analysis of the large $N$ solutions to the BAE \cite{Bobev:2022jte,Bobev:2022eus} and reads
\begin{equation}
\begin{split}
	&\log\Bigg|\fft{\prod_{a=1}^4y_a^{-\fft{N^2}{2}\mn_a}}{\det\mathbb B_\text{TTI}[u_\star,\tu_\star;\Delta]}\fft{\prod_{i=1}^Nx_{\star i}^{N}\tx_{\star i}^{N}\prod_{i\neq j}\left(1-x_{\star i}^{-1}s_{\star j}\right)\left(1-\tx_{\star i}^{-1}\tx_{\star j}\right)}{\prod_{i,j=1}^N\prod_{a=1}^2(\tx_{\star i}-y_ax_{\star j})^{1-\mn_a}\prod_{a=3}^4(x_{\star i}-y_a\tx_{\star j})^{1-\mn_a}}\Bigg|\\
	&=-\fft{\pi\sqrt{2k\Delta_1\Delta_2\Delta_3\Delta_4}}{3}\sum_{a=1}^4\fft{\mn_a}{\Delta_a}\left[\hat N_{k,\Delta}^\fft32-\fft{\mathfrak{c}_a(\Delta)}{k}\hat N_{k,\Delta}^\fft12\right]-\fft12\log\hat N_{k,\Delta}\\
	&\quad+\hat f_0(k,\Delta,\mn)+\hat f_\text{np}(N,k,\Delta,\mn)\,,\label{ABJM:TTI:largeN}
\end{split}
\end{equation}
where $\hat f_\text{np}$ stands for non-perturbative correction of order $\mathcal O(e^{-\sqrt{N}})$ and we have defined
\begin{equation}
	\hat N_{k,\Delta}\equiv N-\fft{k}{24}+\fft{1}{12k}\sum_{a=1}^4\fft{1}{\Delta_a}\,,\qquad \mathfrak c_a(\Delta)\equiv\fft{\prod_{b\neq a}(\Delta_a+\Delta_b)}{8\Delta_1\Delta_2\Delta_3\Delta_4}\sum_{b\neq a}\Delta_a\,.\label{ABJM:TTI:parameters}
\end{equation}
See \cite{Bobev:2022eus} for a discussion on the $N$-independent constant $\hat f_0$ which has no known closed form expression but can be determined accurately with precision numerics. We also emphasize that the all order expression (\ref{ABJM:TTI:largeN}) is given for the specific range of $\Delta$ that satisfies the inequalities (\ref{ABJM:Del:range}) and the constraints (\ref{ABJM:constraint}) as
\begin{equation}
	\sum_{a=1}^4\Delta_a=2\,,\label{ABJM:constraint:Del}
\end{equation}
which we assume from here on. 

We can now use the numerical results for the solution to the BAE from \cite{Bobev:2022jte,Bobev:2022eus} and substitute them in (\ref{ABJM:V}) to deduce a closed form expression for the imaginary part of the Bethe potential that governs the leading term of order $\mathcal O(\omega^{-1})$ in (\ref{ABJM:SCI:Cardy:3}). The result is the following compact expression
\begin{equation}
	\Im\mV_\text{TTI}[u_\star,\tu_\star;\Delta]=2\pi\left[\fft{\pi\sqrt{2k\Delta_1\Delta_2\Delta_3\Delta_4}}{3}\hat N_{k,\Delta}^\fft32+\hat g_0(k,\Delta)+\hat g_\text{np}(N,k,\Delta)\right]\,,\label{ABJM:V:largeN}
\end{equation}
where $\hat g_\text{np}$ stands for non-perturbative correction of order $\mathcal O(e^{-\sqrt{N}})$. We refer the reader to Appendix~\ref{App:ABJM:num} for more details on the numerical analysis by which the closed form expression (\ref{ABJM:V:largeN}) is deduced. It will be very interesting to derive a closed form expression for the $N$-independent constant $\hat g_0$ in \eqref{ABJM:V:largeN}, but we leave this problem for future work.

\medskip

Substituting the subleading term (\ref{ABJM:TTI:largeN}) and the leading term (\ref{ABJM:V:largeN}) back into (\ref{ABJM:SCI:Cardy:3}), we obtain our final result for the all order $1/N$ expansion of the ABJM SCI in the Cardy-like limit
\begin{equation}
	\begin{split}
		&\log\mI_\text{ABJM}(N,k,\omega,\Delta,\mn)\\
		&=-\fft{2}{\omega}\left[\fft{\pi\sqrt{2k\Delta_1\Delta_2\Delta_3\Delta_4}}{3}\hat N_{k,\Delta}^\fft32+\hat g_0(k,\Delta)\right]\\
		&\quad+\left[-\fft{\pi\sqrt{2k\Delta_1\Delta_2\Delta_3\Delta_4}}{3}\sum_{a=1}^4\fft{\mn_a}{\Delta_a}\left(\hat N_{k,\Delta}^\fft32-\fft{\mathfrak{c}_a(\Delta)}{k}\hat N_{k,\Delta}^\fft12\right)-\fft12\log\hat N_{k,\Delta}+\hat f_0(k,\Delta,\mn)\right]\\
		&\quad+\mathcal O(e^{-\sqrt{N}})+\mathcal O(\omega)\,.
	\end{split}\label{ABJM:SCI:Cardy-largeN:1}
\end{equation}
An interesting rewriting of this analytic result for the SCI is possible after first defining new complex parameters $\varphi_a$ as
\begin{equation}
	\varphi_a=\Delta_a+\omega\mn_a~~(e^{\ri\pi\varphi_a}=y_aq^{\mn_a})\quad\to\quad\sum_{a=1}^4\varphi_a=2(1+\omega)\,.\label{ABJM:varphi}
\end{equation}
Using this new variable the ABJM SCI (\ref{ABJM:SCI:Cardy-largeN:1}) can be written more compactly as
\begin{equation}
	\begin{split}
		\log\mI_\text{ABJM}(N,k,\omega,\varphi)&=-\fft43C(k,\omega,\varphi)^{-\fft12}\left(N-B(k,\omega,\varphi)\right)^\fft32-\fft{2}{\omega}\hat g_0(k,\Re\varphi)\\
		&\quad-\fft12\log\left(N-B(k,\omega,\varphi)\right)+\hat f_0(k,\varphi)+\mathcal O(e^{-\sqrt{N}})+\mathcal O(\omega)\,,
	\end{split}\label{ABJM:SCI:Cardy-largeN:2}
\end{equation}
where we have also defined
\begin{subequations}
	\begin{align}
		&C(k,\omega,\varphi)=\fft{2\omega^2}{\pi^2 k\varphi_1\varphi_2\varphi_3\varphi_4}\,,\\
		&B(k,\omega,\varphi)=\fft{k}{24}+\fft{(1+\omega)^2F_1(\varphi)+(1-\omega)^2F_2(\varphi)}{48k\varphi_1\varphi_2\varphi_3\varphi_4}\,,\\
		&F_1(\varphi)=\sum_{a=1}^4\varphi_a^2-\fft{(\varphi_1+\varphi_2-\varphi_3-\varphi_4)(\varphi_1-\varphi_2+\varphi_3-\varphi_4)(\varphi_1-\varphi_2-\varphi_3+\varphi_4)}{\sum_{a=1}^4\varphi_a}\,,\\
		&F_2(\varphi)=-2\sum_{a<b}^4\varphi_a\varphi_b\,.
	\end{align}\label{ABJM:SCI:Cardy-largeN:coeffi}%
\end{subequations}
At leading order in the large $N$ limit, our result \eqref{ABJM:SCI:Cardy-largeN:2} is in agreement with the result obtained recently in \cite{GonzalezLezcano:2022hcf} for the $\mathcal{O}(\omega^{-1})$ and $\mathcal{O}(\omega^0)$ terms in the Cardy-like limit of the ABJM SCI. It would be interesting to understand how to extend the analysis of \cite{GonzalezLezcano:2022hcf} to include finite $N$ corrections. We now proceed to discuss the significance of the rewriting of the leading terms in the Cardy-like limit of the SCI in \eqref{ABJM:SCI:Cardy-largeN:2}.

\subsection{Relation to the Airy conjecture}\label{sec:ABJM:Airy}
Recently, the form of the large $N$ ABJM partition function on various 3d Euclidean supersymmetric backgrounds has been conjectured based on holography and a dual 4d gauged supergravity analysis \cite{Hristov:2021qsw,Hristov:2022lcw}. The ABJM superconformal index, in particular, has been conjecturally given in terms of an Airy function as
\begin{equation}
	\mI_\text{ABJM}^\text{conj}=\left(e^{\mathcal A(k,\omega,\varphi)}C(k,\omega,\varphi)^{-\fft13}\text{Ai}\left[C(k,\omega,\varphi)^{-\fft13}\left(N-B(k,\omega,\varphi)\right)\right]\right)^2(1+\mathcal O(e^{-\sqrt{N}}))\,,\label{ABJM:Airy}
\end{equation}
where the functions $B(k,\omega,\varphi)$ and $C(k,\omega,\varphi)$ are precisely the ones defined in \eqref{ABJM:SCI:Cardy-largeN:coeffi} above. The $N$-independent prefactor $e^{\mathcal A(k,\omega,\varphi)}$ and the non-perturbative correction of order $\mathcal O(e^{-\sqrt{N}})$ have not been conjectured in \cite{Hristov:2021qsw,Hristov:2022lcw}.

The conjecture (\ref{ABJM:Airy}) can be expanded in the Cardy-like limit (\ref{Cardy}) using the asymptotic expansion of the Airy function
\begin{equation}
	\text{Ai}[z]=\fft{e^{-\fft23z^\fft32}}{\sqrt{4\pi}z^\fft14}\left(1+\mathcal O(z^{-\fft32})\right)\,,
\end{equation}
and the result is indeed consistent with the all order result (\ref{ABJM:SCI:Cardy-largeN:2}) we have obtained directly from the matrix model calculation. Importantly, the all order result (\ref{ABJM:SCI:Cardy-largeN:2}) can be used to improve the conjecture by partially restricting the behavior of the unknown prefactor $e^{\mathcal A(k,\omega,\varphi)}$ in the Cardy-like limit to find
\begin{equation}
	\mathcal A(k,\omega,\varphi)=-\fft1\omega\hat g_0(k,\Re\varphi)+\fft{1}{4}\log C(k,\omega,\varphi)+\fft12\hat f_0(k,\varphi)+\fft12\log4\pi+\mathcal O(\omega)\,,
\end{equation}
where the constants $\hat f_0,\hat g_0$ are the ones appearing in (\ref{ABJM:SCI:Cardy-largeN:1}) and can be obtained numerically for various $\varphi$ configurations in \cite{Bobev:2022eus} and Appendix \ref{App:ABJM:num}. Therefore in addition to a non-trivial consistency check of the conjecture in \cite{Hristov:2021qsw,Hristov:2022lcw} for the SCI of the ABJM theory our results can serve as a stepping stone towards the extension of this conjecture to include also $N$-independent contributions.

\section{ADHM superconformal index}\label{sec:ADHM}

In this section we investigate the SCI of the ADHM theory. In Section~\ref{sec:ADHM:mm} we briefly introduce the ADHM theory \cite{Mezei:2013gqa,Grassi:2014vwa} and then provide the matrix model for its SCI based on the general presentation in Section~\ref{sec:general}. We then proceed in Section~\ref{sec:ADHM:Cardy} with the evaluation of the matrix model for the ADHM SCI using the saddle point approximation in the Cardy-like limit following \cite{Choi:2019zpz,Choi:2019dfu}. In Section~\ref{sec:ADHM:largeN} we provide the all order perturbative $1/N$ expansion for the first two leading terms in the Cardy-like expansion of the ADHM SCI using the exact ADHM TTI on $S^1\times S^2$ recently found in \cite{Bobev:2023lkx}.

\subsection{Matrix model for the ADHM SCI}\label{sec:ADHM:mm}
The ADHM theory has a U($N$) gauge group with a vanishing CS level and a corresponding $\mathcal N=4$ vector multiplet, one adjoint $\mathcal N=4$ hypermultiplet, and $N_f$ fundamental $\mathcal N=4$ hypermultiplets. In terms of $\mathcal N=2$ multiplets, the $\mathcal N=4$ vector multiplet decomposes into an $\mathcal N=2$ vector and an $\mathcal N=2$ adjoint chiral multiplet $\Phi_3$ with superconformal $R$-charge 1, the adjoint $\mathcal N=4$ hypermultiplet decomposes into two adjoint $\mathcal N=2$ chiral multiplets $\Phi_{1,2}$ with superconformal $R$-charge $\fft12$, and the $N_f$ fundamental $\mathcal N=4$ hypermultiplets decompose into $N_f$ pairs of fundamental and anti-fundamental $\mathcal N=2$ chiral multiplets $\psi_q$ and $\widetilde\psi_q$ $(q=1,\cdots,N_f)$, respectively, with superconformal $R$-charge $\fft12$. We use $\Phi$ to collectively denote all $\mathcal N=2$ chiral multiplets, i.e. $\Phi=\{\Phi_1,\Phi_2,\Phi_3,\psi_q,\widetilde{\psi}_q\}$.

The global symmetries of the ADHM theory are SU(2)$_\ell\times$SU(2)$_{R_1}\times$SU(2)$_{R_2}\times$U(1)$_T\times {\rm U}(N_f)$. Here SU(2)$_\ell$ corresponds to the flavor symmetry that rotates $\Phi_{1,2}$, SU(2)$_{R_1}\times$SU(2)$_{R_2}$ is the $\mathcal{N}=4$ $R$-symmetry, and U(1)$_T$ is the topological symmetry associated with the U$(N)$ gauge group. The ${\rm U}(N_f)$ flavor symmetry acts only on the $N_f$ fundamental hypermultiplets and will play no role in the discussion below. Let us denote the Cartan generators of the SU(2)$_\ell\times$SU(2)$_{R_1}\times$SU(2)$_{R_2}\times$U(1)$_T$ global symmetries as $(\ell,r_1,r_2,T)$ respectively. Then, from the $\mathcal N=2$ point of view, the superconformal $R$-charge corresponds to $R=r_1+r_2$ and $(\ell,F=r_1-r_2,T)$ become flavor charges \cite{Choi:2019zpz}. The charges of the $\mathcal N=2$ chiral multiplets of the ADHM theory under these symmetries are summarized in Table~\ref{ADHM:global}. 
\begin{table}
	\centering
	\begin{tabular}{|c||c|c|c|c|}
		\hline
		$\Phi$ & $R=r_1+r_2$ & $\ell$ & $F=r_1-r_2$ & $T$ \\\hhline{|=||=|=|=|=|}
		$\Phi_1$ & $\fft12$ & $\fft12$ & $\fft12$ & $0$\\\hline
		$\Phi_2$ & $\fft12$ & $-\fft12$ & $\fft12$ & $0$\\\hline
		$\Phi_3$ & $1$ & $0$ & $-1$ & $0$\\\hline
		$\psi_q$ & $\fft12$ & $0$ & $\fft12$ & $0$\\\hline
		$\widetilde\psi_q$ & $\fft12$ & $0$ & $\fft12$ & $0$\\\hline
	\end{tabular}
	\caption{Charge assignments of the $\mathcal N=2$ chiral multiplets in the ADHM theory. }
	\label{ADHM:global}
\end{table}

The ADHM SCI can be obtained from the trace formula (\ref{SCI:tr:2}) as
\begin{equation}
	\mI_\text{ADHM}(q,\xi_\ell,\xi_F,\xi_T)=\Tr_\text{BPS}\bigg[(-1)^Fq^{R+2j_3}(\xi_\ell)^{\ell}(\xi_F)^{F}(\xi_T)^T\bigg]\,,\label{ADHM:SCI:tr}
\end{equation}
where $(\xi_\ell,\xi_F,\xi_T)$ are fugacities associated with the flavor charges $(\ell,F,T)$ respectively. The corresponding matrix model for the ADHM SCI is then obtained from (\ref{SCI}) using Table~\ref{ADHM:global} and reads
%
\begin{align}\label{ADHM:SCI:1}
	&\mI_\text{ADHM}(q,\xi_\ell,\xi_F,\xi_T)\notag\\
	&=\fft{1}{N!}\sum_{\mm\in\mathbb Z^N}\oint\bigg(\prod_{i=1}^N\fft{dz_i}{2\pi \ri z_i}\xi_T^{\mm_i}\bigg)\prod_{i\neq j}^Nq^{-\fft12|\mm_i-\mm_j|}\left(1-z_iz_j^{-1}q^{|\mm_i-\mm_j|}\right)\notag\\
	&\quad\times\prod_{i,j=1}^N\left(q^{\fft12}z_i^{-1}z_j\xi_\ell^{-\fft12}\xi_F^{-\fft12}\right)^{\fft12|\mm_i-\mm_j|}\fft{(z_i^{-1}z_j\xi_\ell^{-\fft12}\xi_F^{-\fft12}q^{\fft32+|-\mm_i+\mm_j|};q^2)_\infty}{(z_iz_j^{-1}\xi_\ell^\fft12\xi_F^\fft12q^{\fft12+|\mm_i-\mm_j|};q^2)_\infty}\notag\\
	&\quad\times\prod_{i,j=1}^N\left(q^{\fft12}z_i^{-1}z_j\xi_\ell^{\fft12}\xi_F^{-\fft12}\right)^{\fft12|\mm_i-\mm_j|}\fft{(z_i^{-1}z_j\xi_\ell^{\fft12}\xi_F^{-\fft12}q^{\fft32+|-\mm_i+\mm_j|};q^2)_\infty}{(z_iz_j^{-1}\xi_\ell^{-\fft12}\xi_F^\fft12q^{\fft12+|\mm_i-\mm_j|};q^2)_\infty}\\
	&\quad\times\prod_{i,j=1}^N\left(z_i^{-1}z_j\xi_F\right)^{\fft12|\mm_i-\mm_j|}\fft{(z_i^{-1}z_j\xi_Fq^{1+|-\mm_i+\mm_j|};q^2)_\infty}{(z_iz_j^{-1}\xi_F^{-1}q^{1+|\mm_i-\mm_j|};q^2)_\infty}\notag\\
	&\quad\times\prod_{i=1}^N\Bigg[\left(q^{\fft12}z_i^{-1}\xi_F^{-\fft12}\right)^{\fft12|\mm_i|}\fft{(z_i^{-1}\xi_F^{-\fft12}q^{\fft32+|-\mm_i|};q^2)_\infty}{(z_i\xi_F^\fft12q^{\fft12+|\mm_i|};q^2)_\infty}\left(q^{\fft12}z_i\xi_F^{-\fft12}\right)^{\fft12|-\mm_i|}\fft{(z_i\xi_F^{-\fft12}q^{\fft32+|\mm_i|};q^2)_\infty}{(z_i^{-1}\xi_F^\fft12q^{\fft12+|-\mm_i|};q^2)_\infty}\Bigg]^{N_f}\,,\notag
\end{align}
%
where we have turned on the mixed CS term between the gauge symmetry and the topological symmetry only. The phases in the 2nd line of (\ref{SCI}) disappear in the ADHM SCI (\ref{ADHM:SCI:1}) since the phases from adjoint and pairs of fundamental \& anti-fundamental chiral multiplets cancel each other.

As in the ABJM case, to investigate the ADHM SCI in the Cardy-like limit (\ref{Cardy}), it is useful to rewrite it as \cite{Choi:2019zpz,Choi:2019dfu}
\begin{equation}
	\begin{split}
		&\mI_\text{ADHM}(\omega,\Delta,\mn)\\
		&=\fft{1}{N!}\sum_{\mm\in\mathbb Z^N}\oint_{|s_i|=q^{\mm_i}}\bigg(\prod_{i=1}^N\fft{ds_i}{2\pi \ri s_i}(s_i\bs_i)^{-\fft{1}{2\omega}(\Delta_m+\fft{N_f}{2}\Delta_3)+\fft{\mft}{2}-\fft{N_f \mn_3}{4}}\bigg)\\
		&\quad\times(-1)^\fft{N(N-1)}{2}\times\prod_{i=1}^N(s_i\bs_i)^{\fft{N-2i+1}{2}}\times\prod_{i\neq j}^N(1-s_i^{-1}s_j)^\fft12(1-\bs_i^{-1}\bs_j)^\fft12\\
		&\quad\times\prod_{I=1}^3\Bigg[\prod_{i=1}^N\fft{(y_I^{-1}q^{2-\mn_I};q^2)_\infty}{(y_Iq^{\mn_I};q^2)_\infty}\times\prod_{i>j}^N\fft{(s_i^{-1}s_jy_I^{-1}q^{2-\mn_I};q^2)_\infty}{(\bs_i^{-1}\bs_jy_Iq^{\mn_I};q^2)_\infty}\fft{(\bs_i^{-1}\bs_jy_I^{-1}q^{2-\mn_I};q^2)_\infty}{(s_i^{-1}s_jy_Iq^{\mn_I};q^2)_\infty}\Bigg]\\
		&\quad\times\prod_{i=1}^N\Bigg[\fft{(s_i^{-1}y_q^{-1}q^{2-\mn_q};q^2)_\infty}{(\bs_i^{-1}y_qq^{\mn_q};q^2)_\infty}\fft{(\bs_i^{-1}y_{\tq}^{-1}q^{2-\mn_{\tq}};q^2)_\infty}{(s_i^{-1}y_{\tq}q^{\mn_{\tq}};q^2)_\infty}\Bigg]^{N_f}
	\end{split}\label{ADHM:SCI:3}
\end{equation}
in terms of new integration variables (recall that $z_i=e^{\ri h_i},\,q=e^{\ri\pi\omega}$)
\begin{equation}
	s_i=e^{\ri U_i}=z_iq^{\mm_i}=e^{\ri(h_i+\pi\omega\mm_i)}\,,\label{ADHM:s}
\end{equation}
and the parameters $y_I=e^{\ri\pi\Delta_I},y_m=e^{\ri\pi\Delta_m},y_q=e^{\ri\pi\Delta_q},y_{\tq}=e^{\ri\pi\Delta_{\tq}}\,(\Delta_I,\Delta_m,\Delta_q,\Delta_{\tq}\in\mathbb R)$ and $\mn_I,\mft,\mn_q,\mn_{\tq}\in\mathbb R$ defined as
\begin{equation}
	(y_1q^{\mn_1-\fft12},y_2q^{\mn_2-\fft12},y_3q^{\mn_3-1},y_m^{-1}q^{\mft})=(\xi_\ell^\fft12\xi_F^\fft12,\xi_\ell^{-\fft12}\xi_F^\fft12,\xi_F^{-1},\xi_T)\,,\label{ADHM:yn}
\end{equation}
under the constraints
\begin{equation}
	\begin{alignedat}{2}
		\sum_{I=1}^3\mn_I=\mn_3+\mn_q+\mn_{\tq}=2\,,\qquad\prod_{I=1}^3y_I=y_3y_qy_{\tq}=1\,.\label{ADHM:constraint}
	\end{alignedat}
\end{equation}
Note that $(y_I,y_m,\mn_I,\mft)$ are uniquely determined by the three complex parameters $(\xi_\ell,\xi_F,\xi_T)$ through the map (\ref{ADHM:yn}) under the constraints (\ref{ADHM:constraint}). The ADHM SCI does not depend on $\Delta_q,\Delta_{\tq}$ and $\mn_q,\mn_{\tq}$ independently, which have been introduced simply by manipulating the integration variables to make the comparison with the ADHM TTI more straightforward in Section~\ref{sec:ADHM:largeN}. Accordingly in (\ref{ADHM:SCI:3}) we have replaced the argument of the ADHM SCI as
\begin{equation}
	\mI_\text{ADHM}(q,\xi_\ell,\xi_F,\xi_T)\quad\to\quad\mI_\text{ADHM}(\omega,\Delta,\mn)\,,
\end{equation}
where $(\Delta,\mn)$ collectively represent $(\Delta_I,\Delta_m,\mn_I,\mft)$ under the constraints (\ref{ADHM:constraint}). To obtain the expression (\ref{ADHM:SCI:3}) we have also assumed $q\in\mathbb R_{>0}\,(\ri\omega\in\mathbb R)$ as in the ABJM case. More details on the derivation of (\ref{ADHM:SCI:3}) are presented in Appendix~\ref{App:ADHM:Cardy}.

\subsection{Cardy-like limit of the ADHM SCI}\label{sec:ADHM:Cardy}
Taking the Cardy-like limit (\ref{Cardy}), one can simplify the matrix model (\ref{ADHM:SCI:3}) further as in the ABJM case, see \cite{Choi:2019zpz,Choi:2019dfu} and Appendix \ref{App:ADHM:Cardy} for details. The result reads
\begin{equation}
	\begin{split}
		&\mI_\text{ADHM}(\omega,\Delta,\mn)\\
		&=\fft{1}{N!}\int_{\mathbb C^N}\prod_{i=1}^N\fft{dU_id\bU_i}{-4\ri\pi^2\omega}(-1)^{\fft{N(N-1)}{2}}e^{\fft{1}{\pi\omega}\Im\mW^{(0)}[U;\Delta]+2\Re\mW^{(1)}[U;\Delta,\mn]+\mathcal O(\omega)}\,,
	\end{split}\label{ADHM:SCI:Cardy:1}
\end{equation}
where the first two leading terms of the effective action in the Cardy-like limit are given as
\begin{subequations}
	\begin{align}
		\mathcal W^{(0)}[U;\Delta]&=\pi\left(\Delta_m+\fft{N_f}{2}\Delta_3\right)\sum_{i=1}^NU_i+\fft{1}{2}\sum_{I=1}^3\sum_{i=1}^N\bigg[\text{Li}_2(y_I^{-1})-\text{Li}_2(y_I)\bigg]\label{ADHM:W:0}\\
		&\quad+\sum_{I=1}^3\sum_{i>j}^N\bigg[\text{Li}_2(s_i^{-1}s_jy_I^{-1})-\text{Li}_2(s_i^{-1}s_jy_I)\bigg]+N_f\sum_{i=1}^N\bigg[\text{Li}_2(s_i^{-1}y_q^{-1})-\text{Li}_2(s_i^{-1}y_{\tq})\bigg]\,,\nn\\
		\mW^{(1)}[U;\Delta,\mn]&=\fft{\ri}{2}\left(\mft-\fft{N_f}{2}\mn_3+N-2i+1\right)\sum_{i=1}^NU_i+\fft12\sum_{i\neq j}^N\log(1-s_i^{-1}s_j)\nn\\
		&\quad+\fft14\sum_{I=1}^3\sum_{i=1}^N(1-\mn_I)\bigg[\text{Li}_1(y_I^{-1})+\text{Li}_1(y_I)\bigg]\nn\\
		&\quad+\fft12\sum_{I=1}^3\sum_{i>j}^N(1-\mn_I)\bigg[\text{Li}_1(s_i^{-1}s_jy_I^{-1})+\text{Li}_1(s_i^{-1}s_jy_I)\bigg]\nn\\
		&\quad+\fft{N_f}{2}\sum_{i=1}^N\bigg[(1-\mn_{\tq})\text{Li}_1(s_i^{-1}y_{\tq})+(1-\mn_q)\text{Li}_1(s_i^{-1}y_q^{-1})\bigg]\nn\\
		&\quad+\fft{NN_f}{4}\bigg[(1-\mn_{\tq})\log y_{\tq}-(1-\mn_q)\log y_q\bigg]\,.\label{ADHM:W:1}
	\end{align}\label{ADHM:W}%
\end{subequations}
Note that in (\ref{ADHM:SCI:Cardy:1}) we have used the complex conjugation relations
\begin{equation}
	\overline{\mW^{(0)}[U;\Delta]}=\mW^{(0)}[-\bU;-\Delta]\,,\qquad\overline{\mW^{(1)}[U;\Delta,\mn]}=\mW^{(1)}[-\bU;-\Delta,\mn]\,,
\end{equation}
which can be obtained using that $(\Delta,\mn)$ are real by construction, see (\ref{ADHM:yn}).

To apply the saddle point approximation to the integral (\ref{ADHM:SCI:Cardy:1}), first one should solve the saddle point equations in the Cardy-like limit, namely
\begin{equation}
	0=\fft{\partial\mathcal W^{(0)}[U;\Delta]}{\partial U_i}\,.\label{ADHM:Cardy:saddle}
\end{equation}
Let us denote the solution to the leading order saddle point equation (\ref{ADHM:Cardy:saddle}) by $\{U_{\star i}\}$ with the apparent $N!$ degeneracy from permutations. The saddle point approximation to the integral (\ref{ADHM:SCI:Cardy:1}) can then be obtained by following Appendix \ref{App:saddle} to find
\begin{align}\label{ADHM:SCI:Cardy:2}
		&\log\mI_\text{ADHM}(\omega,\Delta,\mn)\notag\\
		&=\fft{1}{\pi\omega}\Im\mW^{(0)}[U_\star;\Delta]\\
		&\quad+\log\Bigg|\fft{\prod_{I=1}^3y_I^{-\fft{N^2}{2}\mn_I}}{\det\mathbb B[U_\star;\Delta]}\fft{\prod_{i=1}^Ns_{\star i}^{N+\mft}\prod_{i\neq j}(1-s_{\star i}^{-1}s_{\star j})}{\prod_{I=1}^3\prod_{i,j=1}^N(s_{\star i}-s_{\star j}y_I)^{1-\mn_I}}\fft{y_{\tq}^{\fft{NN_f(1-\mn_{\tq})}{2}}y_q^{\fft{NN_f(1-\mn_q)}{2}}\prod_{i=1}^Ns_{\star i}^{\fft{N_f}{2}\mn_3}}{\prod_{i=1}^N(1-s_{\star i}y_q)^{N_f(1-\mn_q)}(s_{\star i}-y_{\tq})^{N_f(1-\mn_{\tq})}}\Bigg|\notag\\
		&\quad+\log(-1)^{\fft{N(N-1)}{2}}-\fft12\log(-1)^N+\mathcal O(\omega)\,,\notag
\end{align}
where we have defined the $N\times N$ square matrix $\mathbb B[U;\Delta]$ by its component as
\begin{equation}
	\Big[\mathbb B[U;\Delta]\Big]_{i,j}\equiv\fft{\partial\mW^{(0)}[U;\Delta]}{\partial U_i\partial U_j}\,.\label{ADHM:B:SCI}
\end{equation}
See Appendix \ref{App:ABJM:Cardy} for more details on this derivation.

\subsection{All order $1/N$ expansion for the ADHM SCI in the Cardy-like limit}\label{sec:ADHM:largeN}
To evaluate the Cardy expansion of the ADHM SCI (\ref{ADHM:SCI:Cardy:2}) explicitly in the large $N$ limit, we will use the known all order results for the $S^1\times S^2$ ADHM TTI \cite{Bobev:2022jte,Bobev:2023lkx}. As a first step, we rename the gauge holonomies and the leading order effective action in (\ref{ADHM:SCI:Cardy:2}) as
\begin{equation}
	\begin{split}
		U_i &\to u_i,\qquad \mW^{(0)}[U;\Delta]\to-\mV_\text{TTI}[u;\Delta]\,.\label{ADHM:SCI:to:TTI}
	\end{split}
\end{equation}
Then by using the inversion formula of the polylogarithm (\ref{polylog:inversion}) and assuming
\begin{equation}
	0<\Re[u_i-u_j+\pi\Delta_I]<2\pi\,,\quad -2\pi<\Re[u_i-u_j-\pi\Delta_I]<0\qquad(i<j)\,,\label{ADHM:Del:range}
\end{equation}
one can rewrite $\mV_\text{TTI}[u;\Delta]$ as
\begin{equation}
	\begin{split}
		\mV_\text{TTI}[u;\Delta]&=\pi\left(N-2\ri+1-\Delta_m-\fft{N_f}{2}\Delta_3\right)\sum_{i=1}^Nu_i\\
		&\quad+\fft{1}{2}\sum_{I=1}^3\sum_{i,j=1}^N\bigg[\text{Li}_2(e^{\ri(u_j-u_i+\pi\Delta_I)})-\text{Li}_2(e^{\ri(u_j-u_i-\pi\Delta_I)})\bigg]\\
		&\quad+N_f\sum_{i=1}^N\bigg[\text{Li}_2(e^{\ri(-u_i+\pi\Delta_{\tq})})-\text{Li}_2(e^{\ri(-u_i-\pi\Delta_q)})\bigg]\,.
	\end{split}\label{ADHM:V}
\end{equation}
Similar to the ABJM case, we again find that the function (\ref{ADHM:V}) is exactly the same as the Bethe potential for the $S^1\times S^2$ ADHM TTI~\cite{Bobev:2023lkx}. This implies that the leading order saddle point equations for the matrix model of the ADHM SCI in the Cardy-like limit, (\ref{ADHM:Cardy:saddle}), are equivalent to the Bethe Ansatz Equations (BAE) obtained by taking partial derivatives of the Bethe potential with respect to $u_i$,
\begin{equation}
	0=\fft{\partial\mV_\text{TTI}[u;\Delta]}{\partial u_i}\,.\label{ADHM:TTI:BAE}
\end{equation}
Based on the equivalence between the saddle point equation (\ref{ADHM:Cardy:saddle}) and the BAE (\ref{ADHM:TTI:BAE}) under the map (\ref{ADHM:SCI:to:TTI}), one can rewrite the Cardy expansion of the ADHM SCI (\ref{ADHM:SCI:Cardy:2}) as
\begin{align}\label{ADHM:SCI:Cardy:3}
		&\log\mI_\text{ADHM}(\omega,\Delta,\mn)\notag\\
		&=-\fft{1}{\pi\omega}\Im\mV_\text{TTI}[u_\star;\Delta]\\
		&\quad+\log\Bigg|\fft{\prod_{I=1}^3y_I^{-\fft{N^2}{2}\mn_I}}{\det\mathbb B_\text{TTI}[u_\star;\Delta]}\fft{\prod_{i=1}^Nx_{\star i}^{N+\mft}\prod_{i\neq j}(1-x_{\star i}^{-1}x_{\star j})}{\prod_{I=1}^3\prod_{i,j=1}^N(x_{\star i}-x_{\star j}y_I)^{1-\mn_I}}\fft{y_{\tq}^{\fft{NN_f(1-\mn_{\tq})}{2}}y_q^{\fft{NN_f(1-\mn_q)}{2}}\prod_{i=1}^Nx_{\star i}^{\fft{N_f}{2}\mn_3}}{\prod_{i=1}^N(1-x_{\star i}y_q)^{N_f(1-\mn_q)}(x_{\star i}-y_{\tq})^{N_f(1-\mn_{\tq})}}\Bigg|\notag\\
		&\quad+\fft{iN(N-2)\pi}{2}+\mathcal O(\omega)\,,\notag
\end{align}
in terms of the solutions $\{x_{\star i}=e^{\ri u_{\star i}}\}$ to the BAE (\ref{ADHM:TTI:BAE}) and the $N\times N$ matrix
\begin{equation}
	\Big[\mathbb B_\text{TTI}[u;\Delta]\Big]_{i,j}\equiv-\fft{\partial\mV_\text{TTI}[u;\Delta]}{\partial u_i\partial u_j}\,,\label{ADHM:TTI:B}
\end{equation}
which is equivalent to the Jacobian matrix in the Bethe Ansatz formulation for the ADHM TTI \cite{Hosseini:2016ume,Bobev:2023lkx}. 

Note that the first subleading term of order $\mathcal O(\omega^0)$ in (\ref{ADHM:SCI:Cardy:3}) is precisely the same as the real part of the logarithm of the $S^1\times S^2$ ADHM TTI, which has recently been evaluated to all orders in the perturbative $1/N$ expansion by using the numerical BAE solution \cite{Bobev:2022jte,Bobev:2023lkx} and reads
\begin{align}\label{ADHM:TTI:largeN}
		&\log\Bigg|\fft{\prod_{I=1}^3y_I^{-\fft{N^2}{2}\mn_I}}{\det\mathbb B_\text{TTI}[u_\star;\Delta]}\fft{\prod_{i=1}^Nx_{\star i}^{N+\mft}\prod_{i\neq j}(1-x_{\star i}^{-1}x_{\star j})}{\prod_{I=1}^3\prod_{i,j=1}^N(x_{\star i}-x_{\star j}y_I)^{1-\mn_I}}\fft{y_{\tq}^{\fft{NN_f(1-\mn_{\tq})}{2}}y_q^{\fft{NN_f(1-\mn_q)}{2}}\prod_{i=1}^Nx_{\star i}^{\fft{N_f}{2}\mn_3}}{\prod_{i=1}^N(1-x_{\star i}y_q)^{N_f(1-\mn_q)}(x_{\star i}-y_{\tq})^{N_f(1-\mn_{\tq})}}\Bigg|\notag\\
		&=-\fft{\pi\sqrt{2N_f\tDelta_1\tDelta_2\tDelta_3\tDelta_4}}{3}\sum_{a=1}^4\tmn_a\left[\fft{1}{\tDelta_a}\hat N_{N_f,\tDelta}^\fft32+\left(\mathfrak{c}_a(\tDelta)N_f+\fft{\mathfrak{d}_a(\tDelta)}{N_f}\right)\hat N_{N_f,\tDelta}^\fft12\right]\\
		&\quad-\fft12\log\hat N_{N_f,\tDelta}+\hat f_0(N_f,\tDelta,\tmn)+\hat f_\text{np}(N,N_f,\tDelta,\tmn)\,,\notag
\end{align}
where $\hat f_\text{np}$ stands for non-perturbative corrections of order $\mathcal O(e^{-\sqrt{N}})$ and we have defined
\begin{subequations}
\begin{align}
	\hat N_{N_f,\tDelta}&\equiv N-\fft{N_f}{24}+\fft{N_f}{12}\left(\fft{1}{\tDelta_1}+\fft{1}{\tDelta_2}\right)+\fft{1}{12N_f}\left(\fft{1}{\tDelta_3}+\fft{1}{\tDelta_4}\right)\,,\\
	\mathfrak c_a(\tilde\Delta)&\equiv\bigg(-\fft{1}{\tilde\Delta_1}\fft{(\tilde\Delta_2+\tilde\Delta_3+\tilde\Delta_4)(\tilde\Delta_1+\tilde\Delta_2)}{8\tilde\Delta_1\tilde\Delta_2},-\fft{1}{\tilde\Delta_2}\fft{(\tilde\Delta_1+\tilde\Delta_3+\tilde\Delta_4)(\tilde\Delta_1+\tilde\Delta_2)}{8\tilde\Delta_1\tilde\Delta_2},\nn\\
	&\quad~-\fft{\tilde\Delta_3+\tilde\Delta_4}{8\tilde\Delta_1\tilde\Delta_2},-\fft{\tilde\Delta_3+\tilde\Delta_4}{8\tilde\Delta_1\tilde\Delta_2}\bigg)\,,\\
	\mathfrak d_a(\tilde\Delta)&\equiv\bigg(-\fft{(\tilde\Delta_1+\tilde\Delta_2)(\tilde\Delta_2+\tilde\Delta_3+\tilde\Delta_4)(\tilde\Delta_1+\tilde\Delta_3+\tilde\Delta_4)}{8\tilde\Delta_1\tilde\Delta_2\tilde\Delta_3\tilde\Delta_4},\nn\\
	&\quad~-\fft{(\tilde\Delta_1+\tilde\Delta_2)(\tilde\Delta_2+\tilde\Delta_3+\tilde\Delta_4)(\tilde\Delta_1+\tilde\Delta_3+\tilde\Delta_4)}{8\tilde\Delta_1\tilde\Delta_2\tilde\Delta_3\tilde\Delta_4},\\
	&\quad~-\fft{1}{\tilde\Delta_3}\fft{(\tilde\Delta_3+\tilde\Delta_4)((\tDelta_1+\tDelta_2)(\tDelta_2+\tDelta_3)(\tDelta_3+\tDelta_1)+(\tDelta_1\tDelta_2+\tDelta_2\tDelta_3+\tDelta_3\tDelta_1)\tDelta_4)}{8\tilde\Delta_1\tilde\Delta_2\tilde\Delta_3\tilde\Delta_4},\nn\\
	&\quad~-\fft{1}{\tilde\Delta_4}\fft{(\tilde\Delta_3+\tilde\Delta_4)((\tDelta_1+\tDelta_2)(\tDelta_2+\tDelta_4)(\tDelta_4+\tDelta_1)+(\tDelta_1\tDelta_2+\tDelta_2\tDelta_4+\tDelta_4\tDelta_1)\tDelta_3)}{8\tilde\Delta_1\tilde\Delta_2\tilde\Delta_3\tilde\Delta_4}\bigg)\,,\nn
\end{align}\label{ADHM:TTI:parameters}%
\end{subequations}
in terms of the `tilde' parameters
\begin{subequations}
\begin{align}
	\tilde\Delta&\equiv\bigg(\Delta_1,\Delta_2,\fft{\Delta_3}{2}-\fft{\Delta_m}{N_f},\fft{\Delta_3}{2}+\fft{\Delta_m}{N_f}\bigg)\,,\\
	\tilde\mn&\equiv\bigg(\mn_1,\mn_2,\fft{\mn_3}{2}+\fft{\mathfrak t}{N_f},\fft{\mn_3}{2}-\fft{\mathfrak t}{N_f}\bigg)\,.
\end{align}\label{ADHM:tilde}%
\end{subequations}
We do not know how to derive a closed form expression for the $N$-independent constant $\hat f_0$ but it can be obtained numerically with very good precision, see \cite{Bobev:2023lkx}. We also emphasize that the all order expression (\ref{ADHM:TTI:largeN}) is given for the specific range of $\Delta$ that satisfies the inequalities (\ref{ADHM:Del:range}) and the constraints (\ref{ADHM:constraint}) as
\begin{equation}
	\sum_{a=1}^4\tDelta_a=\sum_{I=1}^3\Delta_I=2\,,\label{ADHM:constraint:Del}
\end{equation}
which we assume from here on.

Substituting exactly the same numerical BAE solutions used to derive the all order ADHM TTI (\ref{ADHM:TTI:largeN}) in \cite{Bobev:2023lkx} into (\ref{ADHM:V}), we find a simple closed form expression for the imaginary part of the Bethe potential that governs the leading term of order $\mathcal O(\omega^{-1})$ in (\ref{ADHM:SCI:Cardy:3}) to find
\begin{equation}
	\Im\mV_\text{TTI}[u_\star;\Delta]=2\pi\left[\fft{\pi\sqrt{2N_f\tDelta_1\tDelta_2\tDelta_3\tDelta_4}}{3}\hat N_{N_f,\Delta}^\fft32+\hat g_0(N_f,\Delta)+\hat g_\text{np}(N,N_f,\Delta)\right]\,,\label{ADHM:V:largeN}
\end{equation}
where $\hat g_\text{np}$ stands for non-perturbative corrections of order $\mathcal O(e^{-\sqrt{N}})$. We present more details on the numerical analysis that allowed us to deduce this expression in Appendix~\ref{App:ADHM:num}. It will be very interesting to find a closed form expression for the $N$-independent constant $\hat g_0$ in \eqref{ADHM:V:largeN}.

\medskip

Substituting the subleading term (\ref{ADHM:TTI:largeN}) and the leading term (\ref{ADHM:V:largeN}) back into (\ref{ADHM:SCI:Cardy:3}), we obtain our final result for the all order $1/N$ expansion of the ADHM SCI in the Cardy-like limit 
\begin{equation}
	\begin{split}
		&\log\mI_\text{ADHM}(N,N_f,\omega,\Delta,\mn)\\
		&=-\fft{2}{\omega}\left[\fft{\pi\sqrt{2N_f\tDelta_1\tDelta_2\tDelta_3\tDelta_4}}{3}\hat N_{N_f,\Delta}^\fft32+\hat g_0(N_f,\Delta)\right]\\
		&\quad-\fft{\pi\sqrt{2N_f\tDelta_1\tDelta_2\tDelta_3\tDelta_4}}{3}\sum_{a=1}^4\tmn_a\left[\fft{1}{\tDelta_a}\hat N_{N_f,\tDelta}^\fft32+\bigg(\mathfrak{c}_a(\tDelta)N_f+\fft{\mathfrak{d}_a(\tDelta)}{N_f}\bigg)\hat N_{N_f,\tDelta}^\fft12\right]\\
		&\quad~~-\fft12\log\hat N_{N_f,\tDelta}+\hat f_0(N_f,\tDelta,\tmn)+\fft{iN(N-2)\pi}{2}+\mathcal O(e^{-\sqrt{N}})+\mathcal O(\omega)\,.
	\end{split}\label{ADHM:SCI:Cardy-largeN:1}
\end{equation}
Following the discussion in the  ABJM case, we can rewrite this result by introducing new complex parameters $\varphi_a$ as
\begin{equation}
	\varphi_a=\tDelta_a+\omega\tmn_a~~(e^{\ri\pi\varphi_a}=\ty_aq^{\tmn_a})\quad\to\quad\sum_{a=1}^4\varphi_a=2(1+\omega)\,.\label{ADHM:varphi}
\end{equation}
After this change of variables the ADHM SCI (\ref{ADHM:SCI:Cardy-largeN:1}) can be written more compactly as
\begin{equation}
	\begin{split}
		&\log\mI_\text{ADHM}(N,N_f,\omega,\varphi)\\
		&=-\fft43C(N_f,\omega,\varphi)^{-\fft12}\left(N-B(N_f,\omega,\varphi)\right)^\fft32-\fft{2}{\omega}\hat g_0(N_f,\Re\varphi)-\fft12\log\left(N-B(N_f,\omega,\varphi)\right)\\
		&\quad+\hat f_0(N_f,\varphi)+\fft{iN(N-2)\pi}{2}+\mathcal O(e^{-\sqrt{N}})+\mathcal O(\omega)\,,
	\end{split}\label{ADHM:SCI:Cardy-largeN:2}
\end{equation}
where we have also defined
\begin{subequations}
	\begin{align}
		C(N_f,\omega,\varphi)&=\fft{2\omega^2}{\pi^2N_f\varphi_1\varphi_2\varphi_3\varphi_4}\,,\\
		B(N_f,\omega,\varphi)&=\fft{N_f}{24}-\fft{N_f}{12}\left(\fft{1}{\varphi_1}+\fft{1}{\varphi_2}\right)-\fft{1}{12N_f}\left(\fft{1}{\varphi_3}+\fft{1}{\varphi_4}\right)\nn\\
		&\quad+\omega\left(\fft{N_f}{12}\fft{\varphi_3+\varphi_4}{\varphi_1\varphi_2}+\fft{(\varphi_1+\varphi_2)(\varphi_1+\varphi_3+\varphi_4)(\varphi_2+\varphi_3+\varphi_4)}{12N_f\varphi_1\varphi_2\varphi_3\varphi_4}\right)\,.
	\end{align}\label{ADHM:SCI:Cardy-largeN:coeffi}%
\end{subequations}
It will be interesting to use these SCI results, together with the TTI results in \cite{Bobev:2022jte,Bobev:2023lkx}, to obtain a conjecture similar to the one proposed in \cite{Hristov:2021qsw,Hristov:2022lcw} for the large $N$ partition function of the ADHM theory on general 3d Euclidean supersymmetric backgrounds. 

\section{Holography and black holes}\label{sec:holo}

In this section we discuss the holographic implications of the ABJM and ADHM SCI results derived above. First we focus on the so-called universal KN AdS$_4$ black hole solution which arises in special limits of the parameters and can be obtained in 4d minimal $\mathcal N=2$ gauged supergravity. We then proceed to discuss the more general case of supersymmetric KN AdS$_4$ black hole solutions in 4d $\mathcal N=2$ gauged supergravity coupled to vector multiplets.

\subsection{Holography in the universal limit}\label{sec:holo:mini}

The universal limit of an Euclidean SCFT partition function, as discussed in \cite{Azzurli:2017kxo,Bobev:2017uzs,Bobev:2019zmz}, amounts to turning on fugacities and background fields associated only with the energy momentum tensor multiplet of the SCFT. When viewed as 3d $\mathcal{N}=2$ SCFTs the ABJM and ADHM theories admit such a treatment which amounts to the following special values of the parameters used above
\begin{equation}
	\text{ABJM :}~~\Delta_a=\mn_a=\fft12\,,\qquad\text{ADHM :}~~\tDelta_a=\tmn_a=\fft12\,.
\end{equation}
In this universal limit the general expressions (\ref{ABJM:SCI:Cardy-largeN:1}) and (\ref{ADHM:SCI:Cardy-largeN:1}) reduce to
\begin{subequations}
	\begin{align}
		&\log\mI_\text{ABJM}(N,k,\omega)\nn\\
		&=-\fft{\pi\sqrt{2k}}{3}\Bigg[\left(\frac{1}{2\omega}+1\right)\left(N-\fft{k}{24}+\fft{2}{3k}\right)^\fft32-\fft3k\left(N-\fft{k}{24}+\fft{2}{3k}\right)^\fft12\Bigg]\nn\\
		&\quad-\fft{2}{\omega}\hat g_0(k)-\fft12\log\left(N-\fft{k}{24}+\fft{2}{3k}\right)+\hat f_0(k)+\mathcal O(e^{-\sqrt{N}})+\mathcal O(\omega)\,,\label{ABJM:SCI:Cardy-largeN:special}\\
		&\log\mI_\text{ADHM}(N,N_f,\omega)\nn\\
		&=-\fft{\pi\sqrt{2N_f}}{3}\Bigg[\left(\frac{1}{2\omega}+1\right)\left(N+\fft{7N_f}{24}+\fft{1}{3N_f}\right)^\fft32-\left(\fft{N_f}{2}+\fft{5}{2N_f}\right)\left(N+\fft{7N_f}{24}+\fft{1}{3N_f}\right)^\fft12\Bigg]\nn\\
		&\quad-\fft{2}{\omega}\hat g_0(N_f)-\fft12\log\left(N+\fft{7N_f}{24}+\fft{1}{3N_f}\right)+\hat f_0(N_f)+\mathcal O(e^{-\sqrt{N}})+\mathcal O(\omega)\,,\label{ADHM:SCI:Cardy-largeN:special}
	\end{align}\label{SCI:Cardy-largeN:special}%
\end{subequations}
respectively. 

The holographic dual of this universal limit of the SCI is given by the Euclidean supersymmetric KN AdS$_4$ black hole solution of the $\mathcal N=2$ gauged minimal supergravity \cite{Kostelecky:1995ei,Caldarelli:1998hg} uplifted to 11d through the consistent truncation on $S^7$ \cite{deWit:1986oxb,Gauntlett:2007ma}. The difference in the supergravity description between the ABJM and ADHM models originates in the different orbifold action resulting in the distinct internal manifolds $S^7/\mathbb Z_k$ and $S^7/\mathbb Z_{N_f}$, see \cite{Bobev:2021oku} for more details. The large $N$ QFT results in (\ref{ABJM:SCI:Cardy-largeN:special}) and (\ref{ADHM:SCI:Cardy-largeN:special}) can be viewed as field theory predictions for the Euclidean path integral of M-theory on these two asymptotically AdS$_4$ Euclidean backgrounds. The first two leading terms of order $N^\fft32$ and $N^\fft12$ in the logarithm of the Euclidean M-theory path integral can been obtained by evaluating (minus) the regularized Euclidean on-shell action of the dual supersymmetric KN AdS$_4$ black hole solution. The $N^\fft32$ term arises from the 2-derivative on-shell action whereas the $N^\fft12$ term can be accounted for by using four-derivative corrections to the minimal 4d $\mathcal{N}=2$ gauged supergravity, see \cite{Cassani:2019mms,Bobev:2020egg,Bobev:2021oku}. Using the results of \cite{Hristov:2021zai} one can also calculate the $\log N$ contribution to the SCI from the dual supergravity description. Combining these supergravity results we find the following holographic expectation for the form of the SCI of the ABJM and ADHM models in the large $N$ limit\footnote{In (\ref{M-theory:Z}) the subscript ``f'' in the product symbol should remind the reader that the 11d solution involves a non-trivial fibration between the external and the internal manifolds.}
\begin{subequations}
\begin{align}
	&\log Z_\text{M-theory}\big|_{\text{KN\,AdS}_4\times_\text{f}S^7/\mathbb Z_k}\nn\\
	&=-\fft{\pi\sqrt{2k}}{3}\Bigg[\fft{(\omega+1)^2}{2\omega}\bigg(N^\fft32+\bigg(\fft{1}{k}-\fft{k}{16}\bigg)N^\fft12\bigg)-\fft3kN^\fft12\Bigg]-\frac{1}{2}\log N +\ldots\,,\\
	&\log Z_\text{M-theory}\big|_{\text{KN\,AdS}_4\times_\text{f}S^7/\mathbb Z_{N_f}}\nn\\
	&=-\fft{\pi\sqrt{2N_f}}{3}\Bigg[\fft{(\omega+1)^2}{2\omega}\bigg(N^\fft32+\bigg(\fft{1}{2N_f}+\fft{7N_f}{16}\bigg)N^\fft12\bigg)-\fft{N_f^2+5}{2N_f}N^\fft12\Bigg]-\frac{1}{2}\log N +\ldots\,.
\end{align}\label{M-theory:Z}%
\end{subequations}
It is now easy to check that the all order $1/N$ expansion for the ABJM and ADHM SCI (\ref{SCI:Cardy-largeN:special}) in the Cardy-like limit is indeed identical to the logarithm of the dual Euclidean M-theory path integral (\ref{M-theory:Z}) up to and including the terms of order  $\log N$ and $\omega^0$. This amounts to a highly non-trivial precision test of holography and improves on previous results in the literature where this agreement was demonstrated in the leading $N^\fft32\omega^{-1}$ order \cite{Choi:2019zpz}.

Note that the field theory SCI result in (\ref{SCI:Cardy-largeN:special}) does not have precise information about subleading corrections of order $\mathcal O(\omega)$ in the Cardy-like limit but still provides the all order prediction in the $1/N$ expansion for the dual Euclidean M-theory path integral for the $\omega^{-1}$ and $\omega^0$ terms. On the other hand, even though only the first three leading terms of order $N^\fft32$, $N^\fft12$, and $\log N$ have been obtained from supergravity in (\ref{M-theory:Z}), their exact $\omega$-dependence is fully determined and can be used to restrict the subleading corrections of order $\mathcal O(\omega)$ in the dual SCI (\ref{SCI:Cardy-largeN:special}). In particular, this leads to a supergravity prediction about the $N^\fft32$ and $N^\fft12$ terms at order $\omega$ (and their absence at higher orders in $\omega$) as well as the absence of any $\log N$ term at order $\omega$ or higher in the small $\omega$ limit. It will be most interesting to confirm these supergravity results independently by field theory calculations.

Another important holographic application of the SCI (\ref{SCI:Cardy-largeN:special}) is that the Legendre transform of the SCI determines the entropy function of the dual supersymmetric KN AdS$_4$ black hole in the large $N$ limit \cite{Choi:2018fdc,Cassani:2019mms,Hristov:2019mqp,Bobev:2019zmz}. To see this explicitly for the minimal supergravity of interest here, let us write down the Legendre transform of the SCI (\ref{SCI:Cardy-largeN:special}) in the large $N$ limit as\footnote{Compared to the convention of \cite{Bobev:2019zmz}, we have $(\omega,\varphi)^\text{there}=2\pi \ri(\omega,\varphi)^\text{here}$ and restore the AdS$_4$ radius $L$ explicitly. In \cite{Bobev:2019zmz} the angular momentum $J$ is dimensionless but the electric charge $Q$ has inverse length dimension so we put the extra factor of $L$ in front of $Q$ in (\ref{entropy:fct}).}
\begin{subequations}
	\begin{align}
		\mathcal{S}_\text{ABJM}(N,k,\omega,\varphi,J,Q,\lambda)&=-\fft{2\pi\sqrt{2k}}{3}\fft{\varphi^2}{\omega}N^\fft32-2\pi \ri(\omega J+\varphi LQ+\lambda(2\varphi-1-\omega))\,,\\
		\mathcal{S}_\text{ADHM}(N,N_f,\omega,\varphi,J,Q,\lambda)&=-\fft{2\pi\sqrt{2N_f}}{3}\fft{\varphi^2}{\omega}N^\fft32-2\pi \ri(\omega J+\varphi LQ+\lambda(2\varphi-1-\omega))\,,
	\end{align}\label{entropy:fct}%
\end{subequations}
where the parameters $\omega$ and $\varphi$ play the roles of chemical potentials associated with the black hole angular momentum $J$ and electric charge $Q$, respectively, and we have also introduced the Lagrange multiplier $\lambda$ to implement the constraint $2\varphi=1+\omega$. Extremizing the entropy functions (\ref{entropy:fct}) with respect to $\omega,\varphi,\lambda$ and demanding positive entropy at the extremum then gives the Bekenstein-Hawking entropy of a dual supersymmetric KN AdS$_4$ black hole as discussed in \cite{Choi:2018fdc,Cassani:2019mms,Hristov:2019mqp,Bobev:2019zmz}
\begin{equation}
		S_{\text{KN\,AdS}_4}(J(Q),Q)=\fft{\pi L}{G_N}\fft{J(Q)}{Q}=\fft{\pi L^2}{2G_N}\Bigg[\sqrt{1+\fft{4G_N^2Q^2}{L^2}}-1\Bigg]\,.\label{entropy}
\end{equation}
Note that the 4d Newton constant $G_N$ is related to the field theory parameters in the large $N$ limit as (see \cite{Bobev:2021oku} for instance)
\begin{equation}
	\fft{L^2}{2G_N}=\begin{cases}
		\fft{\sqrt{2k}}{3}N^\fft32 & (\text{for ABJM})\\
		\fft{\sqrt{2N_f}}{3}N^\fft32 & (\text{for ADHM})
	\end{cases}\,.\label{GL:NkNf}
\end{equation}
Note also that in (\ref{entropy}), the constraint $J = \frac{LQ}{2}\Big[\sqrt{1+4G_N^2Q^2/L^2}-1\Big]$ between the angular momentum $J$ and the electric charge $Q$ of the Lorentzian black hole does not arise from the extremization procedure but follows from imposing a reality condition on the entropy. We discuss this reality constraint further below.

It is worth emphasizing that the entropy function (\ref{entropy:fct}) cannot be obtained relying solely on the all order $1/N$ expansion of the SCI (\ref{SCI:Cardy-largeN:special}). Since the latter is obtained in the Cardy-like limit up to subleading corrections of order $\mathcal O(\omega)$, the all order SCI (\ref{SCI:Cardy-largeN:special}) does not provide enough information to determine the leading term in the entropy function (\ref{entropy:fct}) unambiguously to take the form $\sim\fft{\varphi^2}{\omega}N^\fft32$ with the constraint $2\varphi=1+\omega$. Hence we have implicitly used the known result for the holographic dual on-shell action (\ref{M-theory:Z}) obtained from supergravity, which is not exact in the $1/N$ expansion but exact for generic finite $\omega$, to fully determine the $N^\fft32$ term in the entropy function in (\ref{entropy:fct}). It will be very interesting to employ a field theory calculation of the ABJM and ADHM SCI for generic finite $\omega$ to rigorously derive the large $N$ entropy function in (\ref{entropy:fct}).

If the all order $1/N$ expansion of the ABJM and ADHM SCI in the Cardy-like limit (\ref{SCI:Cardy-largeN:special}) can be found for finite $\omega$, one can use it to derive the entropy of the dual supersymmetric KN AdS$_4$ black hole beyond the large $N$ limit. To be precise, we expect that the latter can be obtained from the former by implementing the inverse Laplace transform as
\begin{equation}
\begin{split}
	\Omega(J,Q)&=\int d\lambda\int_0^1d\omega d\varphi\,\mI_\text{ABJM/ADHM}(\omega,\varphi)e^{-2\pi \ri(\omega J+\varphi LQ+\lambda(2\varphi-1-\omega))}\,,\\
	S_{\text{KN\,AdS}_4}(J(Q),Q)&=\log\Omega(J,Q)\Big|_{\text{Fix }J=J(Q)\text{ by demanding real }\log\Omega(J,Q)}\,.\label{entropy:exact}
\end{split}
\end{equation}
Note that the entropy function (\ref{entropy:fct}) can be read off from the exponent of the integrand in (\ref{entropy:exact}) by taking the large $N$ limit of the ABJM or ADHM SCI as in (\ref{SCI:Cardy-largeN:special}). The Bekenstein-Hawking entropy (\ref{entropy}) then arises from the large $N$ saddle point approximation of the integral in (\ref{entropy:exact}). To derive corrections to the Bekenstein-Hawking formula one needs to evaluate the integral (\ref{entropy:exact}) beyond the large $N$ limit. This is an interesting and important open question for future research. Its answer necessitates an expression for the large $N$ expansion of the ABJM/ADHM SCI valid for generic finite $\omega$ beyond the Cardy-like limit we studied here.

It is worth discussing further the positive entropy condition, which has been used to obtain the dual supersymmetric KN AdS$_4$ black hole entropy (\ref{entropy}) by extremizing the entropy function (\ref{entropy:fct}) read off from the SCI (\ref{SCI:Cardy-largeN:special}). We are not aware of a field theory argument for imposing such a positivity condition. More specifically, it is not a priori justified to impose positivity on the logarithm of the microcanonical partition function obtained from the SCI via the inverse Laplace transform in (\ref{entropy:exact}) especially since we have already allowed for general complex chemical potentials. The positive entropy condition arises naturally only if we interpret the logarithm of the microcanonical partition function as the entropy of a holographically dual KN AdS$_4$ black hole. This implies that the microcanonical partition function $\Omega(J,Q)$ given in (\ref{entropy:exact}) requires an appropriate holographic dual interpretation before imposing positivity on $\log\Omega(J,Q)$. It is important to understand whether a purely field theoretic argument can be found that imposes such a positivity constraint on the SCI of holographic SCFTs. This puzzle is alleviated in Euclidean signature where one can interpret the generic complex $\log\Omega(J,Q)$ as the ``entropy'' associated with a dual supersymmetric Euclidean supergravity solution satisfying the canonical thermodynamic relation \cite{Cassani:2019mms,Bobev:2019zmz}. In this sense, the positive entropy condition arises only if one analytically continues these supergravity solutions to Lorentzian signature and wants to interpret them as regular supersymmetric KN AdS$_4$ black hole. In view of this subtlety, we consider the Euclidean M-theory path integral discussed around \eqref{M-theory:Z} as the more natural setting to study precision holography for the SCI.

\subsection{Holography for general charges}\label{sec:holo:general}
The holographic dual description of the ABJM and ADHM SCI given in (\ref{ABJM:SCI:Cardy-largeN:2}) and (\ref{ADHM:SCI:Cardy-largeN:2}) is expected to be the supersymmetric KN AdS$_4$ black hole with generic values of the electric charges obtained as a solution in $\mathcal N=2$ gauged supergravity coupled to vector multiplets with a certain prepotential. For both the ABJM and ADHM model, a dual KN AdS$_4$ black hole solution with two different electric charges in the $\mathcal N=2$ gauged supergravity coupled to a single vector multiplet (the $X^0X^1$ model) was found in \cite{Cvetic:2005zi}.\footnote{The reason for this result is that this supergravity model is simply the 4d $\mathcal{N}=4$ SO$(4)$ gauged supergravity which should be the common universal holographic sector shared by the ABJM and ADHM models according to the proposal in \cite{Bobev:2017uzs}.}  The dual description of the ABJM SCI with generic fugacities, is in terms of  the dyonic KN AdS$_4$ black hole solutions in the $\mathcal N=2$ gauged supergravity coupled to three vector multiplets (the STU model) recently constructed in \cite{Hristov:2019mqp}. From here on for concreteness we focus our brief discussion on the ABJM SCI (\ref{ABJM:SCI:Cardy-largeN:2}) with generic configurations for $\varphi$ as in (\ref{ABJM:varphi}).

As in the minimal supergravity case discussed in Section~\ref{sec:holo:mini}, the ABJM SCI (\ref{ABJM:SCI:Cardy-largeN:2}) with generic $\varphi$ is expected to describe  the M-theory path integral around a Euclidean supersymmetric KN AdS$_4$ black hole solution in the STU model uplifted to 11d through the consistent truncation on $S^7$ derived in \cite{Cvetic:1999xp,Azizi:2016noi}. The leading $N^\fft32$ term in the SCI should correspond to minus the regularized Euclidean on-shell action of the dual supersymmetric KN AdS$_4$ solution. To the best of our knowledge this has not been explicitly verified for general values of the fugacities. For the special case with two different electric charges in the STU model (which amounts to studying the $X^0X^1$ model) this calculation of the on-shell action has been performed in detail in \cite{Cassani:2019mms} yielding the result\footnote{Compared to the convention of \cite{Cassani:2019mms}, we have $(\omega,\varphi_a)^\text{there}=2\pi \ri(\omega,\varphi_a)^\text{here}$, $4Q_a^\text{there}=Q_a^\text{here}$, and we restore the AdS$_4$ radius $L$ and the 4d Newton constant $G_N$ explicitly.}
\begin{equation}
\begin{split}
	\log Z_\text{M-theory}\big|_{\text{KN\,AdS}_4\times_\text{f}S^7/\mathbb Z_k}=-\fft{2\pi\sqrt{2k}}{3}\fft{\varphi_1\varphi_3}{\omega}N^\fft32+\mathcal O(N^\fft12)\,.
\end{split}
\end{equation}
Note that here we made use of the map between the gravitational and field theory parameters in (\ref{GL:NkNf}). This results agrees with the logarithm of the ABJM SCI (\ref{ABJM:SCI:Cardy-largeN:2}) at the $N^\fft32$ leading order upon the identification $\varphi_1=\varphi_2$ and $\varphi_3=\varphi_4$. Note again that the ABJM SCI (\ref{ABJM:SCI:Cardy-largeN:2}) is derived in the Cardy-like limit of small $\omega$ whereas the supergravity result is valid for general finite $\omega$. This is analogous to the minimal gauged supergravity discussion in Section~\ref{sec:holo:mini}. The $N^\fft12$ term in the large $N$ expansion of the ABJM SCI (\ref{ABJM:SCI:Cardy-largeN:2}) has not yet been derived using supergravity methods. The expectation is that it can be accounted for by studying the four-derivative generalization of the STU model of 4d $\mathcal{N}=2$ gauged supergravity. A discussion of this model was recently presented in \cite{Bobev:2021oku}, however, it has not yet been applied in the context of holography for rotating charged black holes. It will of course be very interesting to pursue this problem further. The $\log N$ term in the ABJM SCI (\ref{ABJM:SCI:Cardy-largeN:2}) on the other hand can be computed using supergravity. This has been discussed in some detail in \cite{Hristov:2021zai} where it was shown that the coefficient of this term is independent of the fugacities and always takes the form $-\frac{1}{2}\log N$ in the large $N$ limit. This agrees with the field theory result in (\ref{ABJM:SCI:Cardy-largeN:2}) providing yet another test of holography and solidifying further the analysis in \cite{Hristov:2021zai}.

The analysis in \cite{Choi:2019zpz} of the ABJM SCI in the large $N$ limit taken after the Cardy-like limit leads to the following proposal for the entropy function for the supersymmetric KN AdS$_4$ black hole in the STU gauged supergravity model 
\begin{equation}
\begin{split}
	\mathcal{S}_\text{ABJM}(N,k,\omega,\varphi,J,Q,\lambda)&=-\fft{2\pi\sqrt{2k}}{3}\fft{\sqrt{\varphi_1\varphi_2\varphi_3\varphi_4}}{\omega}N^\fft32\\
	&\quad-2\pi \ri\bigg(\omega J+\fft{L}{4}\sum_{a=1}^4\varphi_aQ_a+\lambda\bigg(\fft12\sum_{a=1}^4\varphi_a-1-\omega\bigg)\bigg)\,.
\end{split}\label{entropy:fct:general}
\end{equation}
This proposal for the entropy function has been tested by matching it with the Bekenstein-Hawking entropy of the known black hole solutions in \cite{Cvetic:2005zi} upon extremization with respect to $\omega,\varphi,\lambda$ and after imposing a reality condition on the entropy \cite{Choi:2018fdc,Cassani:2019mms}. A further generalization of this to an entropy function for the dyonic black holes in the STU model was presented in~\cite{Hristov:2019mqp}. Note that the entropy function (\ref{entropy:fct:general}) reduces to the minimal one (\ref{entropy:fct}) for identical electric charges $Q_a$ and associated chemical potentials $\varphi_a$. However, as discussed in Section~\ref{sec:holo:mini}, the derivation of the entropy function (\ref{entropy:fct:general}) from the ABJM SCI is still incomplete since the logarithm of the ABJM SCI is obtained only up to $\mathcal O(\omega)$ in (\ref{ABJM:SCI:Cardy-largeN:1}) and therefore its $N^\fft32$ leading order does not completely fix the dependence of the entropy function in (\ref{entropy:fct:general}) on the fugacities. This once again underscores the importance of extending our result for the ABJM SCI (\ref{ABJM:SCI:Cardy-largeN:1}) beyond the Cardy-like limit and finding the index for general finite values of the fugacities.

\section{Discussion}\label{sec:discussion}

The central result in this work is the explicit evaluation of the leading $\omega^{-1}$ and the first subleading $\omega^0$ order in the Cardy-like limit of small $\omega$ for the SCI of the ABJM and ADHM 3d holographic SCFTs. Our calculations are based on the observation that these two terms in the SCI are closely related to the Bethe potential and the TTI of the SCFT.\footnote{This relation between the SCI and TTI at order $\omega^{-1}$ can be explained by using the decomposition of these two partition functions into the Cardy blocks discussed in \cite{Choi:2019dfu}. We believe that there is a similar explanation for the relation between the SCI and TTI at order $\omega^0$ which we plan to explore in future work.} These two quantities can in turn be computed to all orders in the perturbative $1/N$ expansion using high-precision numerical calculations based on the recent results in \cite{Bobev:2022jte,Bobev:2022eus,Bobev:2023lkx}. This analysis amounts to the closed form expressions in \eqref{ABJM:SCI:Cardy-largeN:1} and \eqref{ADHM:SCI:Cardy-largeN:1} for the two leading terms in the Cardy-like limit of the SCI. In the large $N$ limit these results can be successfully compared to holographically dual calculations performed using 2- and 4-derivative supergravity as well as 1-loop corrections to supergravity. We also discussed the implications of these results to the microscopic entropy counting of the dual supersymmetric charged and rotating AdS$_4$ black hole solutions. Our analysis points to several interesting questions for future studies which we now briefly discuss.

A natural generalization of our results would be to study the SCI of more general 3d $\mathcal N=2$ holographic SCFTs in the Cardy-like limit. In particular, it is important to establish whether for such models the first two leading terms in the small $\omega$ limit of the SCI can in general be identified with the Bethe potential and the TTI of the corresponding SCFT. If this is indeed the case then one can use the recent TTI results \cite{Bobev:2023lkx} valid to all order in the $1/N$ expansion of such holographic SCFTs to compute the corresponding SCI for these models. We hope to investigate this question in the near future.

Another interesting extension of our work is to study the SCI in the presence of background magnetic fluxes for continuous flavor symmetries. Since the flavor magnetic fluxes can be turned on simply by shifting the gauge magnetic fluxes appropriately \cite{Kapustin:2011jm}, deriving the formal expression for the Cardy-like expansion of this generalized SCI in terms of the saddle point approximation would be analogous to what we did in Section~\ref{sec:ABJM} and Section~\ref{sec:ADHM} above. The non-trivial question is to understand whether the map between the SCI and the Bethe potential and TTI continues to hold in some form in the presence of these flavor magnetic fluxes. If indeed such a map exists then it can be leveraged to derive the all order $1/N$ expansion of the generalized SCI in the Cardy-like limit which in turn can be used to explore the physics of the dual dyonic KN AdS$_4$ black holes. This question is especially relevant for holographic applications to the ABJM theory where such dyonic black hole solutions in the STU model of 4d gauged supergravity were recently studied in \cite{Hristov:2019mqp}.

As we emphasized throughout this paper the all order $1/N$ expansion of the SCI we have obtained for the ABJM and ADHM SCI is restricted to the first two leading terms of order $\omega^{-1}$ and $\omega^0$ in the Cardy-like limit. We were not able to study the subleading corrections of order $\mathcal O(\omega)$ in the SCI relying on the saddle point approximation. New calculational methods may be needed to find such corrections. It will be most interesting to understand how to calculate the large $N$ SCI for holographic SCFTs without appealing to the small $\omega$ limit. As discussed in Section~\ref{sec:holo}, this is particularly important in the context of holography since the fugacities and charges of the dual supergravity solutions are in general finite. For the ABJM theory, in particular, finding the SCI in the large $N$ limit at finite $\omega$  can also be used to test the Airy conjecture of \cite{Hristov:2021qsw,Hristov:2022lcw} beyond the leading two orders in the small $\omega$ expansion which we confirmed in this work.

Our focus here was on the M-theory limit of the SCI for the ABJM  and ADHM models. This corresponds to taking $N$ large and keeping $k$ and $N_f$ fixed. The recent results in \cite{Bobev:2022eus} strongly suggest that the all order $1/N$ expansion results for the SCI in \eqref{ABJM:SCI:Cardy-largeN:1} and \eqref{ADHM:SCI:Cardy-largeN:1} can be reorganized to yield the type IIA expansion characterized by large $N$ and $k$ (or $N_f$) with fixed $k/N$ (or $N_f/N$). It will be interesting to analyze this in more detail and in particular to understand whether one can use string theory techniques to access the $1/N$ corrections to the SCI from the holographically dual type IIA description.

Given the results presented above for the SCI, as well as analogous explicit results for the all order $1/N$ expansion of the $S^3$ and TTI partition functions of holographic SCFTs, see \cite{Marino:2016new,Bobev:2022jte,Bobev:2022eus,Bobev:2023lkx}, it is natural to wonder whether similar exact results can be derived for the large $N$ expansion of the partition functions of these holographic models on more general compact Euclidean manifolds of the type discussed in \cite{Closset:2019hyt}. It will be most interesting to pursue this question and understand the general lessons that can be drawn from it in order to further increase our understanding of precision holography, black hole physics, and quantum gravity.

\section*{Acknowledgements}

We are grateful to Francesco Benini, Anthony Charles, Shai Chester, Fri\dh rik Freyr Gautason, Yasuyuki Hatsuda, Kiril Hristov, Chiung Hwang, Marcos Mari\~no, Dario Martelli, Leo Pando-Zayas, Silviu Pufu, Yu Xin, and Alberto Zaffaroni for valuable discussions. NB and JH are supported in part by an Odysseus grant G0F9516N from the FWO. NB, JH, and VR are also supported by the KU Leuven C1 grant ZKD1118 C16/16/005. SC is supported by a KIAS Individual Grant (PG081601) at the Korea Institute for Advanced Study. SC and JH are grateful to Seoul National University for hospitality during the early gestation stage of this project. VR would like to thank IPhT Saclay for its hospitality during the final stages of the preparation of the manuscript.

\appendix

\section{Special functions}\label{App:special}

The polylogarithm is defined within the unit disk as
\begin{equation}
	\text{Li}_n(z)\equiv\sum_{k=1}^\infty\fft{z^k}{k^n}\qquad(|z|<1)\,,
\end{equation}
and then extended to $|z|\geq1$ by analytic continuation with the branch cut. The polylogarithm has the following inversion formula
\begin{equation}
	\text{Li}_n(e^{\ri u})+(-1)^n\text{Li}_n(e^{-\ri u})=-\fft{(2\pi \ri)^n}{n!}B_n\Big(\fft{u}{2\pi}\Big)~~\text{for}~~\begin{cases}
	0\leq\Re[u]<2\pi~\&~\Im[u]\geq0\\
	0<\Re[u]\leq 2\pi~\&~\Im[u]<0 
	\end{cases}\,,\label{polylog:inversion}
\end{equation}
in terms of the Bernoulli polynomials $B_n(x)$.

The $\infty$-Pochhammer symbol is defined within the unit disk as
\begin{equation}
	(a;q)_\infty=\prod_{n=0}^\infty(1-aq^n)\qquad(|q|<1)\,,\label{eq:poch}
\end{equation}
and can be extended to $|q|>1$, see Appendix A of \cite{Choi:2019dfu} for more details. The $\infty$-Pochhammer symbol satisfies the identity
\begin{equation}
	(-x)^\fft{\mm}{2}\fft{(xq^{1+\mm};q^2)_\infty}{(x^{-1}q^{1+\mm};q^2)_\infty}=(-x)^{-\fft{\mm}{2}}\fft{(xq^{1-\mm};q^2)_\infty}{(x^{-1}q^{1-\mm};q^2)_\infty}\qquad(\mm\in\mathbb Z)\,.\label{poch:inverse}
\end{equation}
The $\infty$-Pochhammer symbol also has the following asymptotic expansion
\begin{equation}
\begin{split}
	\lim_{|q|\to1^-}(aq^m;q^2)_\infty&=\exp[-\fft{\ri}{2\pi\omega}\text{Li}_2(aq^{m-1})](1+\mathcal O(\omega))\quad(a\in\mathbb C,~a\notin[1,\infty),~q=e^{\ri\pi\omega})\\
	&=\exp[-\fft{\ri}{2\pi\omega}\text{Li}_2(a)+\fft{m-1}{2}\text{Li}_1(a)](1+\mathcal O(\omega))
\end{split}\label{poch:asymp}
\end{equation}
in terms of the polylogarithm functions, see Appendix A of \cite{Choi:2019zpz,Choi:2019dfu} for more details.

\section{Saddle point approximation}\label{App:saddle}
Here we consider the saddle point approximation for the following type of integral,
\begin{equation}
	Z(\omega)=\int\prod_{i=1}^n\fft{du_i}{2\pi}\exp[-\fft{\ri}{2\pi\omega}\left(\mF^{(0)}[u]+2\ri\pi\omega\mF^{(1)}[u]+\mathcal O(\omega^2)\right)]\,,\label{Z:1}
\end{equation}
in the $\omega\to \ri0^+$ limit. To begin with, we find a saddle point perturbatively
\begin{equation}
	u_\star=\sum_{k=0}^\infty u_\star^{(k)}\omega^k\,,\label{u:star}
\end{equation}
where the first two coefficients in the small $\omega$-expansion are determined by solving the saddle point equations perturbatively as
\begin{subequations}
	\begin{align}
		0&=\fft{\partial}{\partial u_i}\mF^{(0)}[u]\bigg|_{u=u_\star^{(0)}}\,,\label{Cardy:saddle:lead}\\
		0&=\fft{\partial}{\partial u_i}\left(\mF^{(0)}+2\ri\pi\omega\mF^{(1)}\right)[u]\bigg|_{u=u_\star^{(0)}+\omega u_\star^{(1)}}+\mathcal O(\omega^2)\,.\label{Cardy:saddle:sublead}
	\end{align}\label{Cardy:saddle}%
\end{subequations}
Let us denote the degeneracy of the saddle point (\ref{u:star}) with $d_n$. The saddle point approximation for the integral (\ref{Z:1}) then reads
\begin{equation}
	\begin{split}
		&Z(\omega)\\
		&=d_n\exp[-\fft{\ri}{2\pi\omega}\left(\mF^{(0)}[u_\star^{(0)}+\omega u_\star^{(1)}]+2\ri\pi\omega\mF^{(1)}[u_\star^{(0)}]+\mathcal O(\omega^2)\right)]\\
		&\quad\times\int\prod_{i=1}^n\fft{(-\ri\pi\omega)^\fft12d\Delta u_i}{2\pi}e^{-\fft12\bigg(\fft{1}{2!}\fft{\partial^2\mF^{(0)}[u_\star^{(0)}]}{\partial u_i\partial u_j}\Delta u_i\Delta u_j+\fft{(-\ri\pi\omega)^\fft12}{3!}\fft{\partial^3\mF^{(0)}[u_\star^{(0)}]}{\partial u_i\partial u_j\partial u_k}\Delta u_i\Delta u_j\Delta u_k+\mathcal O(\omega)\bigg)}\\
		&=d_n\exp[-\fft{\ri}{2\pi\omega}\mF^{(0)}[u_\star^{(0)}+\omega u_\star^{(1)}]+\mF^{(1)}[u_\star^{(0)}]]\left(-\ri\omega\right)^{\fft{n}{2}}\left(\det\fft{\partial^2\mF^{(0)}[u_\star^{(0)}]}{\partial u_i\partial u_j}\right)^{-\fft12}(1+\mathcal O(\omega))\,,
	\end{split}\label{Z:2}
\end{equation}
where we have introduced the integration variable $\Delta u_i$ as $(-\ri\pi\omega)^\fft12\Delta u_i=(u-u_\star^{(0)}-\omega u_\star^{(1)})_i$ and also used the property
\begin{equation}
	\int\prod_{i=1}^ndx_i\exp[-\sum_{i=1}^NA_ix_i^2]x_jx_kx_l=0\quad~(\text{for } \forall j,k,l\in\{1,\cdots,N\})\,,
\end{equation}
which follows from the fact that the integrand is an odd function. Taking the logarithm of (\ref{Z:2}) and using the leading order saddle point equation (\ref{Cardy:saddle:lead}), we obtain
\begin{equation}
	\begin{split}
		\log Z(\omega)&=-\fft{\ri}{2\pi\omega}\mF^{(0)}[u_\star^{(0)}]+\mF^{(1)}[u_\star^{(0)}]+\fft{n}{2}\log(-\ri\omega)-\fft12\log\det\fft{\partial^2\mF^{(0)}[u_\star^{(0)}]}{\partial u_i\partial u_j}\\
		&\quad+\log d_n+\mathcal O(\omega)\,.
	\end{split}\label{Z:3}
\end{equation}
This is the saddle point approximation used to obtain the ABJM SCI (\ref{ABJM:SCI:Cardy:2}) and the ADHM SCI (\ref{ADHM:SCI:Cardy:2}) in the Cardy-like limit (\ref{Cardy}).

\section{ABJM superconformal index}\label{App:ABJM}

In this Appendix we provide the key intermediate steps that we have skipped in the main text to derive the all order perturbative $1/N$ expansion for the ABJM SCI in the Cardy-like limit~(\ref{ABJM:SCI:Cardy-largeN:1}). In Section~\ref{App:ABJM:Cardy} we present some of the details on the rewriting of the matrix model~(\ref{ABJM:SCI:1}) into the form suitable for the Cardy expansion~(\ref{ABJM:SCI:Cardy:2}) following \cite{Choi:2019zpz,Choi:2019dfu}. In Section~\ref{App:ABJM:num} we discuss the numerical analysis used to evaluate the leading term of order $\mathcal O(\omega^{-1})$ in the Cardy expansion (\ref{ABJM:SCI:Cardy:3}), which is equivalent to (\ref{ABJM:SCI:Cardy:2}), to all orders in the perturbative $1/N$ expansion.

\subsection{Cardy-like expansion of the ABJM SCI}\label{App:ABJM:Cardy}

The first step is to rewrite the matrix model for the ABJM SCI (\ref{ABJM:SCI:1}) as (\ref{ABJM:SCI:3}). We start by shifting the integration variables of the matrix model (\ref{ABJM:SCI:1}) as $(z_i,\tz_i)\to(z_i\xi_{T_1}^{-1/k},\tz_i\xi_{T_2}^{1/k})$, upon which the ABJM SCI (\ref{ABJM:SCI:1}) reads
\begin{equation}
	\begin{split}
		&\mI_\text{ABJM}(\omega,\Delta,\mn)\\
		&=\fft{1}{(N!)^2}\sum_{\mm,\tmm\in\mathbb Z^N}\oint\bigg(\prod_{i=1}^N\fft{dz_i}{2\pi \ri z_i}\fft{d\tz_i}{2\pi \ri\tz_i}z_i^{ k\mm_i}\tz_i^{-k\tmm_i}\bigg)\times\prod_{i\neq j}^N(-1)^{\fft12|\mm_i-\mm_j|+\fft12|\tmm_i-\tmm_j|}\\
		&\quad\times\prod_{i\neq j}^N(-z_iz_j^{-1}q)^{-\fft12|\mm_i-\mm_j|}\fft{(z_i^{-1}z_jq^{|\mm_i-\mm_j|};q^2)_\infty}{(z_iz_j^{-1}q^{2+|\mm_i-\mm_j|};q^2)_\infty}(-\tz_i\tz_j^{-1}q)^{-\fft12|\tmm_i-\tmm_j|}\fft{(\tz_i^{-1}\tz_jq^{|\tmm_i-\tmm_j|};q^2)_\infty}{(\tz_i\tz_j^{-1}q^{2+|\tmm_i-\tmm_j|};q^2)_\infty}\\
		&\quad\times\prod_{a=1}^2\prod_{i,j=1}^N\left(-q^{1-\mn_a}z_i^{-1}\tz_jy_a^{-1}\right)^{\fft12|\mm_i-\tmm_j|}\fft{(z_i^{-1}\tz_jy_a^{-1}q^{2-\mn_a+|\mm_i-\tmm_j|};q^2)_\infty}{(z_i\tz_j^{-1}y_aq^{\mn_a+|\mm_i-\tmm_j|};q^2)_\infty}\\
		&\quad\times\prod_{a=3}^4\prod_{i,j=1}^N\left(-q^{1-\mn_a}z_i\tz_j^{-1}y_a^{-1}\right)^{\fft12|\mm_i-\tmm_j|}\fft{(z_i\tz_j^{-1}y_a^{-1}q^{2-\mn_a+|\mm_i-\tmm_j|};q^2)_\infty}{(z_i^{-1}\tz_jy_aq^{\mn_a+|\mm_i-\tmm_j|};q^2)_\infty}\,
	\end{split}\label{ABJM:SCI:2}
\end{equation}
in terms of the parameters introduced in (\ref{ABJM:yn}). Then we remove the absolute signs for gauge magnetic fluxes in (\ref{ABJM:SCI:2}) as \cite{Choi:2019zpz,Choi:2019dfu}
\begin{equation}
	\begin{split}
		|\mm_i-\tmm_j|&\to\begin{cases}
			-\mm_i+\tmm_j & (i\geq j)\\
			\mm_i-\tmm_j & (i<j)
		\end{cases}\,,\\
		|\mm_i-\mm_j|&\to\begin{cases}
			-(\mm_i-\mm_j) & (i>j)\\
			\mm_i-\mm_j & (i<j)
		\end{cases}\,,\\
		|\tmm_i-\tmm_j|&\to\begin{cases}
			-(\tmm_i-\tmm_j) & (i>j)\\
			\tmm_i-\tmm_j & (i<j)
		\end{cases}\,,
	\end{split}
\end{equation}
using the identity (\ref{poch:inverse}) and reorganize the integrand in terms of the new integration variables introduced in (\ref{ABJM:s}). The result of this procedure is given in (\ref{ABJM:SCI:3}).\footnote{To obtain (\ref{ABJM:SCI:3}) from (\ref{ABJM:SCI:2}) we used $\prod_{a=1}^4(y_a^{-1})^{\fft12|\mm_i-\tmm_j|}=1$ based on (\ref{ABJM:constraint}), which involves a subtle phase issue due to fractional exponents. We plan to explore this issue more thoroughly in future work.}

\medskip

Next, one can expand the expression (\ref{ABJM:SCI:3}) in the Cardy-like limit (\ref{Cardy}) using the asymptotic formula of the $\infty$-Pochhammer symbol (\ref{poch:asymp}) as
\begin{equation}
\begin{split}
	&\mI_\text{ABJM}(\omega,\Delta,\mn)\\
	&=\fft{1}{(N!)^2}\sum_{\mm,\tmm\in\mathbb Z^N}\int_{|s_i|=q^{\mm_i},|\ts_i|=q^{\tmm_i}}\bigg(\prod_{i=1}^N\fft{ds_i}{2\pi \ri s_i}\fft{d\ts_i}{2\pi \ri\ts_i}\bigg)\,e^{\fft{1}{\pi\omega}\Im\mW^{(0)}[U,\tU;\Delta]+2\Re\mW^{(1)}[U,\tU;\Delta,\mn]+\mathcal O(\omega)}\,,
\end{split}\label{ABJM:SCI:Cardy:1-1}
\end{equation}
where the effective actions are defined in (\ref{ABJM:W}). To obtain the expression (\ref{ABJM:SCI:Cardy:1}) from (\ref{ABJM:SCI:Cardy:1-1}), one should replace the discrete sums over gauge magnetic fluxes with the continuous integrals using the Euler-Maclaurin formula, see for example \cite{Pasquetti:2019uop,Choi:2019zpz,Choi:2019dfu}
\begin{equation}
	\sum_{\mm,\tmm\in\mathbb Z^N}\oint_{|s_i|=q^{\mm_i},|\ts_i|=q^{\tmm_i}}\prod_{i=1}^N\fft{ds_i}{2\pi \ri s_i}\fft{d\ts_i}{2\pi \ri\ts_i}(\cdots)=\int_{\mathbb C^{2N}}\prod_{i=1}^N\fft{dU_id\bU_i}{-4\ri\pi^2\omega}\fft{d\tU_id\btU_i}{-4\ri\pi^2\omega}(\cdots)\left(1+\mathcal O(\omega)\right)\,.\label{sum:to:integral}
\end{equation}
To show explicitly how (\ref{sum:to:integral}) works, here we focus on the 1-dimensional integral:
\begin{equation}
	\begin{split}
		&\sum_{\mm\in\mathbb Z}\oint_{|s|=e^{\ri\pi\omega\mm}}\fft{ds}{2\pi \ri s}f(s)=\lim_{M\to\infty}\sum_{\mm=-M}^M\int_0^{2\pi}\fft{d\theta}{2\pi}\,f(e^{\ri\pi\omega\mm+\ri\theta})\\
		&=\int_{-\infty}^\infty d\mm\int_0^{2\pi}\fft{d\theta}{2\pi}\,f(e^{\ri\pi\omega\mm+\ri\theta})\\
		&\quad+\underbrace{\fft{f(0)+f(\infty)}{2}+\sum_{k=1}^{\lfloor\fft{p}{2}\rfloor}\fft{B_{2k}}{(2k)!}(\ri\pi\omega)^{2k-1}\left(s\fft{d}{ds}\right)^{2k-1}f(s)\bigg|^{s=0}_{s=\infty}+\lim_{M\to\infty}\int_0^{2\pi}\fft{d\theta}{2\pi}R_p}_{=\mathcal O(\omega^0)}\\
		&=\int_{\mathbb C}\fft{dUd\bU}{-4\ri\pi^2\omega}\,f(s)\left(1+\mathcal O(\omega)\right)\,,
	\end{split}\label{sum:to:integral:1d}
\end{equation}
where in the second equation we have used the Euler-Maclaurin formula
\begin{equation}
	\begin{split}
		\sum_{i=m}^nF(i)&=\int_m^ndi\,F(i)+\fft{F(n)+F(m)}{2}+\sum_{i=1}^{\lfloor\fft{p}{2}\rfloor}\fft{B_{2k}}{(2k)!}\left(F^{(2k-1)}(n)-F^{2k-1}(m)\right)+R_p\,,\\
		|R_p|&\leq\fft{2\zeta(p)}{(2\pi)^p}\int_m^ndx\,|F^{(p)}(x)|\,,
	\end{split}
\end{equation}
and in the third equation we used the Jacobian determinant ($\ri U=\ri\pi\omega\mm+\ri\theta$)
\begin{equation}
	dUd\bU=\left|\fft{\partial(U,\bU)}{\partial(\mm,\theta)}\right|d\mm d\theta=\left|\begin{pmatrix}
		\pi\omega & 1\\
		-\pi\omega & 1
	\end{pmatrix}\right|d\mm d\theta=-2\ri\pi\omega d\mm d\theta\,.
\end{equation}
Recall that the Cardy-like limit (\ref{Cardy}) is taken as $\omega\to i0^+$ where $-i\omega>0$. The crucial assumption in (\ref{sum:to:integral:1d}) is that the integrand $f(s)$ and its logarithmic derivatives behave nicely in the $s\to0,\infty$ limit as
\begin{equation}
	f(s)~\&~\left(s\fft{d}{ds}\right)^{2k-1}f(s)<\infty~~(k\in\mathbb Z)\qquad\text{as}\qquad s\to0,\infty
\end{equation}
and therefore can be estimated to be of order $\mathcal O(\omega^0)$. In this paper we assume that the integrand of the ABJM SCI (\ref{ABJM:SCI:Cardy:1-1}) satisfies the analogous boundary conditions for the multi-dimensional version of the 1d integral in \eqref{sum:to:integral:1d} and thereby the replacement of the discrete sum over gauge magnetic fluxes with the continuous integrals, (\ref{sum:to:integral}), is valid. It will be interesting to carefully study the validity of this assumption.

\medskip

The last step to obtain the Cardy expansion of the ABJM SCI (\ref{ABJM:SCI:Cardy:2}) is to apply the saddle point approximation introduced in Appendix \ref{App:saddle} to the integral (\ref{ABJM:SCI:Cardy:1}). This can be done by mapping the parameters in Appendix \ref{App:saddle} to the those for the ABJM SCI in (\ref{ABJM:SCI:Cardy:1}) as
\begin{equation}
\begin{split}
	u_i\quad&\to\quad U_i,\bU_i,\tU_i,\btU_i\quad(n=4N)\,,\\
	d_n\quad&\to\quad (N!)^2\,,\\
	\mathcal F^{(0)}[u]\quad&\to\quad2\ri\Im\mW^{(0)}[U,\tU;\Delta]\,,\\
	\mathcal F^{(1)}[u]\quad&\to\quad2\Re\mW^{(1)}[U,\tU;\Delta,\mn]\,,
\end{split}\label{ABJM:saddle}
\end{equation}
and then simply applying the approximation (\ref{Z:3}).

\subsection{Numerical analysis for the ABJM SCI in the Cardy-like limit}\label{App:ABJM:num}
To evaluate the leading term of order $\mathcal O(\omega^{-1})$ in the Cardy expansion of the ABJM SCI (\ref{ABJM:SCI:Cardy:3}), which is given in terms of the Bethe potential (\ref{ABJM:V}) evaluated at the solutions to the BAE (\ref{ABJM:TTI:BAE}), first we construct numerical solutions to the BAE (\ref{ABJM:TTI:BAE}) for $N=101\sim301$ in step of 10 with
\begin{equation}
\begin{split}
	k&\in\{1,2,3,4\}\,,\\
	\Delta&\in\left\{(\fft12,\fft12,\fft12,\fft12),(\fft37,\fft12,\fft12,\fft47),(\fft13,\fft{5}{12},\fft{7}{12},\fft23),(\fft1\pi,\fft2\pi,\fft{3}{2\pi},2-\fft{9}{2\pi})\right\}\,,
\end{split}\label{ABJM:configurations}
\end{equation}
using \emph{Mathematica}, see \cite{Bobev:2022eus} for more details on the numerical procedure. Then we use \texttt{LinearModelFit} to determine the exact values of the Bethe potential (\ref{ABJM:V}) evaluated at the numerical BAE solutions with two fitting functions in terms of the `shifted' $N$ parameter, $\hat N_{k,\Delta}$ in (\ref{ABJM:TTI:parameters}), as
\begin{equation}
	\fft{1}{2\pi}\Im\mV_\text{TTI}=\hat g_{3/2}^\text{(lmf)}(k,\Delta)\hat N_{k,\Delta}^\fft32+\hat g_0^\text{(lmf)}(k,\Delta)\,,
\end{equation}
where the superscript ``(lmf)'' means that the coefficients are determined numerically for a given configuration of $(k,\Delta)$ via \texttt{LinearModelFit}. We then confirm that the leading coefficient $\hat g_{3/2}^\text{(lmf)}(k,\Delta)$ matches the analytic expression
\begin{equation}
	\hat g_{3/2}(k,\Delta)=\fft{\pi\sqrt{2k\Delta_1\Delta_2\Delta_3\Delta_4}}{3}\,,
\end{equation}
read off from (\ref{ABJM:V:largeN}) with great precision for all configurations listed in (\ref{ABJM:configurations}). For a precise comparison, below we provide tables for the error ratio
\begin{equation}
	R_{3/2}(k,\Delta)\equiv\fft{\hat g_{3/2}^\text{(lmf)}(k,\Delta)-\hat g_{3/2}(k,\Delta)}{\hat g_{3/2}(k,\Delta)}\,,
\end{equation}
and the numerical estimate for the constant term $\hat g_0^\text{(lmf)}(k,\Delta)$ together with the associated standard error $\sigma_0$.

\medskip

\noindent$\boldsymbol{\Delta=(\fft12,\fft12,\fft12,\fft12)}$
%
\begin{center}
	\footnotesize
	\begin{tabular}{ |c||c|c|c| } 
		\hline
		& $R_{3/2}$ & $\hat g_0^\text{(lmf)}$ & $\sigma_0$\\
		\hline\hline
		$k=1$ & $-2.198{\times}10^{-42}$ & $-0.18384102333840097854$ & $1.656{\times}10^{-39}$ \\
		\hline
		$k=2$ & $-1.272{\times}10^{-30}$ & $-0.15224228529196635390$ & $1.300{\times}10^{-27}$\\
		\hline
		$k=3$ & $-4.479{\times}10^{-26}$ & $-0.16038269262168390318$ & $5.370{\times}10^{-23}$\\ 
		\hline
		$k=4$ & $-6.404{\times}10^{-23}$ & $-0.18269074235035962450$ & $8.518{\times}10^{-20}$\\ 
		\hline
	\end{tabular}
\end{center}
%

\noindent$\boldsymbol{\Delta=(\fft37,\fft12,\fft12,\fft47)}$
%
\begin{center}
	\footnotesize
	\begin{tabular}{ |c||c|c|c| } 
		\hline
		& $R_{3/2}$ & $\hat g_0^\text{(lmf)}$ & $\sigma_0$\\
		\hline\hline
		$k=1$ & $-5.100{\times}10^{-38}$ & $-0.18487381932915327786$ & $3.757{\times}10^{-35}$ \\
		\hline
		$k=2$ & $-1.134{\times}10^{-27}$ & $-0.15225543589256328299$ & $1.116{\times}10^{-24}$\\
		\hline
		$k=3$ & $-3.618{\times}10^{-24}$ & $-0.15976369368639724921$ & $4.153{\times}10^{-21}$\\ 
		\hline
		$k=4$ & $-2.252{\times}10^{-21}$ & $-0.18141737511951941056$ & $2.851{\times}10^{-18}$\\ 
		\hline
	\end{tabular}
\end{center}
%

\noindent$\boldsymbol{\Delta=(\fft13,\fft{5}{12},\fft{7}{12},\fft23)}$
%
\begin{center}
	\footnotesize
	\begin{tabular}{ |c||c|c|c| } 
		\hline
		& $R_{3/2}$ & $\hat g_0^\text{(lmf)}$ & $\sigma_0$\\
		\hline\hline
		$k=1$ & $-1.179{\times}10^{-29}$ & $-0.19164058508102652614$ & $7.700{\times}10^{-27}$ \\
		\hline
		$k=2$ & $-2.312{\times}10^{-21}$ & $-0.15267798840052025763$ & $1.929{\times}10^{-18}$\\
		\hline
		$k=3$ & $1.751{\times}10^{-20}$ & $-0.15546742123991081702$ & $1.652{\times}10^{-17}$\\ 
		\hline
		$k=4$ & $3.084{\times}10^{-18}$ & $-0.17182133959082622268$ & $3.149{\times}10^{-15}$\\ 
		\hline
	\end{tabular}
\end{center}
%

\noindent$\boldsymbol{\Delta=(\fft1\pi,\fft2\pi,\fft{3}{2\pi},2-\fft{9}{2\pi})}$
%
\begin{center}
	\footnotesize
	\begin{tabular}{ |c||c|c|c| } 
		\hline
		& $R_{3/2}$ & $\hat g_0^\text{(lmf)}$ & $\sigma_0$\\
		\hline\hline
		$k=1$ & $-5.526{\times}10^{-31}$ & $-0.19164115564794187181$ & $3.703{\times}10^{-28}$ \\
		\hline
		$k=2$ & $-1.584{\times}10^{-22}$ & $-0.15308092766631389200$ & $1.370{\times}10^{-19}$\\
		\hline
		$k=3$ & $-1.971{\times}10^{-19}$ & $-0.15795628160679798011$ & $1.940{\times}10^{-16}$\\ 
		\hline
		$k=4$ & $-2.338{\times}10^{-17}$ & $-0.17749207056697949250$ & $2.500{\times}10^{-14}$\\ 
		\hline
	\end{tabular}
\end{center}
%

\section{ADHM superconformal index}\label{App:ADHM}
In this Appendix we provide the key intermediate steps that we have skipped in the main text to derive the all order perturbative $1/N$ expansion for the ADHM SCI in the Cardy-like limit (\ref{ADHM:SCI:Cardy-largeN:1}). In Section~\ref{App:ADHM:Cardy} we provide some details on how to go from the matrix model in (\ref{ADHM:SCI:1}) to the Cardy expansion (\ref{ADHM:SCI:Cardy:2}) following \cite{Choi:2019zpz,Choi:2019dfu}. In Section~\ref{App:ADHM:num} we discuss the numerical analysis used to evaluate the leading term of order $\mathcal O(\omega^{-1})$ in the Cardy expansion (\ref{ADHM:SCI:Cardy:3}), which is equivalent to (\ref{ADHM:SCI:Cardy:2}), to all orders in the perturbative $1/N$ expansion.

\subsection{Cardy-like expansion of the ADHM SCI}\label{App:ADHM:Cardy}
The first step is to rewrite the matrix model for the ADHM SCI (\ref{ADHM:SCI:1}) as (\ref{ADHM:SCI:3}). We start by shifting the integration variables as $z_i\to z_i\xi_F^{-1/2}y_qq^{\mn_q-1/2}$, upon which the ADHM SCI (\ref{ADHM:SCI:1}) reads
\begin{equation}
	\begin{split}
		&\mI_\text{ADHM}(\omega,\Delta,\mn)\\
		&=\fft{1}{N!}\sum_{\mm\in\mathbb Z^N}\oint\bigg(\prod_{i=1}^N\fft{dz_i}{2\pi \ri z_i}(y_m^{-1}q^{\mft})^{\mm_i}\bigg)\times\prod_{i=1}^N(-1)^{-N_f|\mm_i|}\\
		&\quad\times\prod_{i\neq j}^N(-z_iz_j^{-1}q)^{-\fft12|\mm_i-\mm_j|}\fft{(z_i^{-1}z_jq^{|\mm_i-\mm_j|};q^2)_\infty}{(z_iz_j^{-1}q^{2+|\mm_i-\mm_j|};q^2)_\infty}\\
		&\quad\times\prod_{I=1}^3\prod_{i,j=1}^N\left(-q^{1-\mn_I}z_i^{-1}z_jy_I^{-1}\right)^{\fft12|\mm_i-\mm_j|}\fft{(z_i^{-1}z_jy_I^{-1}q^{2-\mn_I+|\mm_i-\mm_j|};q^2)_\infty}{(z_iz_j^{-1}y_Iq^{\mn_I+|\mm_i-\mm_j|};q^2)_\infty}\\
		&\quad\times\prod_{i=1}^N\Bigg[\left(-q^{1-\mn_q}z_i^{-1}y_q^{-1}\right)^{\fft12|\mm_i|}\fft{(z_i^{-1}y_q^{-1}q^{2-\mn_q+|\mm_i|};q^2)_\infty}{(z_iy_qq^{\mn_q+|\mm_i|};q^2)_\infty}\\
		&\kern4em\times\left(-q^{1-\mn_{\tq}}z_iy_{\tq}^{-1}\right)^{\fft12|\mm_i|}\fft{(z_iy_{\tq}^{-1}q^{2-\mn_{\tq}+|\mm_i|};q^2)_\infty}{(z_i^{-1}y_{\tq}q^{\mn_{\tq}+|\mm_i|};q^2)_\infty}\Bigg]^{N_f}\,,
	\end{split}\label{ADHM:SCI:2}
\end{equation}
in terms of the parameters introduced in (\ref{ADHM:yn}). Then we remove the absolute signs for gauge magnetic fluxes in (\ref{ADHM:SCI:2}) by using \cite{Choi:2019zpz,Choi:2019dfu}
\begin{equation}
	\begin{split}
		|\mm_i-\mm_j|&\to\begin{cases}
			-\mm_i+\mm_j & (i\geq j)\\
			\mm_i-\mm_j & (i<j)
		\end{cases}\,,\\
		|\mm_i|&\to -\mm_i\,.
	\end{split}
\end{equation}
We then use the identity (\ref{poch:inverse}) and reorganize the integrand in terms of the new integration variables introduced in (\ref{ADHM:s}). The result of this procedure is given in (\ref{ADHM:SCI:3})

\medskip

The next step is to expand the expression (\ref{ADHM:SCI:3}) in the Cardy-like limit using the asymptotic formula of the $\infty$-Pochhammer symbol (\ref{poch:asymp}) and then replace the discrete sums over gauge magnetic fluxes with the continuous integrals by using \cite{Pasquetti:2019uop,Choi:2019zpz,Choi:2019dfu}
\begin{equation}
	\sum_{\mm\in\mathbb Z^N}\oint_{|s_i|=q^{\mm_i}}\prod_{i=1}^N\fft{ds_i}{2\pi \ri s_i}(\cdots)=\int_{\mathbb C^{N}}\prod_{i=1}^N\fft{dU_id\bU_i}{-4\ri\pi^2\omega}(\cdots)\left(1+\mathcal O(\omega)\right)\,.
\end{equation}
We then use this to finally obtain (\ref{ADHM:SCI:Cardy:1}). This procedure is exactly parallel to the ABJM case reviewed in Appendix~\ref{App:ABJM:Cardy} so we skip the details here.

\medskip

The last step to obtain the Cardy expansion of the ADHM SCI (\ref{ADHM:SCI:Cardy:2}) is to apply the saddle point approximation introduced in Appendix \ref{App:saddle} to the integral (\ref{ADHM:SCI:Cardy:1}). This can be done by mapping the parameters in Appendix \ref{App:saddle} to the those for the ADHM SCI in (\ref{ADHM:SCI:Cardy:1}) as
\begin{equation}
	\begin{split}
		u_i\quad&\to\quad U_i,\bU_i \quad(n=2N)\,,\\
		d_n\quad&\to\quad N!\,,\\
		\mathcal F^{(0)}[u]\quad&\to\quad2\ri\Im\mW^{(0)}[U;\Delta]\,,\\
		\mathcal F^{(1)}[u]\quad&\to\quad2\Re\mW^{(1)}[U;\Delta,\mn]\,,
	\end{split}\label{ADHM:saddle}
\end{equation}
and then simply applying the approximation (\ref{Z:3}).

\subsection{Numerical analysis for the ADHM SCI in the Cardy-like limit}\label{App:ADHM:num}
To evaluate the leading term of order $\mathcal O(\omega^{-1})$ in the Cardy expansion of the ADHM SCI (\ref{ADHM:SCI:Cardy:3}), which is given in terms of the Bethe potential (\ref{ADHM:V}) evaluated at the solutions to the BAE (\ref{ADHM:TTI:BAE}), first we construct numerical solutions to the BAE (\ref{ADHM:TTI:BAE}) for $N=101\sim301$ in step of 10 with
\begin{equation}
	\begin{alignedat}{4}
		&1)&\quad N_f&\in\{1,2,3,4\}\,,&\quad\Delta_I&=(\fft12,\fft12,1,0)\,,&\quad (\Delta_{\tq},\Delta_q)&=(\fft12,\fft12)\,,\\
		&2)&\quad N_f&\in\{1,2,3\}\,,&\quad\Delta_I&=(\fft37,\fft47,1,0)\,,&\quad (\Delta_{\tq},\Delta_q)&=(\fft12,\fft12)\,,\\
		&3)&\quad N_f&\in\{1,2,3\}\,,&\quad\Delta_I&=(\fft38,\fft58,1,\fft{N_f}{10})\,,&\quad (\Delta_{\tq},\Delta_q)&=(\fft12,\fft12)\,,\\
		&\text{4-i})&\quad N_f&\in\{1,2,3\}\,,&\quad\Delta_I&=(\fft1\pi,\fft2\pi,2-\fft3\pi,N_f\left(1-\fft3\pi\right))\,,&\quad (\Delta_{\tq},\Delta_q)&=(\fft{3}{2\pi},\fft{3}{2\pi})\,,\\
		&\text{4-ii})&\quad N_f&\in\{3,4\}\,,&\quad\Delta_I&=(\fft1\pi,\fft2\pi,2-\fft3\pi,N_f\left(1-\fft3\pi\right))\,,&\quad (\Delta_{\tq},\Delta_q)&=(\fft{e}{2\pi},\fft{3}{\pi}-\fft{e}{2\pi})\,,
	\end{alignedat}\label{ADHM:configurations}
\end{equation}
using \emph{Mathematica}, see \cite{Bobev:2023lkx} for more details on the numerical procedure. Recall that $\Delta$ denotes $\Delta=(\Delta_I,\Delta_m)$ for the ADHM case. Then we use \texttt{LinearModelFit} to determine the exact values of the Bethe potential (\ref{ABJM:V}) evaluated at the numerical BAE solutions with two fitting functions in terms of the `shifted' $N$ parameter, $\hat N_{N_f,\tDelta}$ in (\ref{ADHM:TTI:parameters}), as
\begin{equation}
	\fft{1}{2\pi}\Im\mV_\text{TTI}=\hat g_{3/2}^\text{(lmf)}(N_f,\Delta)\hat N_{N_f,\tDelta}^\fft32+\hat g_0^\text{(lmf)}(N_f,\Delta)\,,
\end{equation}
where the superscript ``(lmf)'' means that the coefficients are determined numerically for a given configuration of $(N_f,\Delta)$ via \texttt{LinearModelFit}. We then confirm that the leading coefficient $\hat g_{3/2}^\text{(lmf)}(N_f,\Delta)$ matches the analytic expression
\begin{equation}
	\hat g_{3/2}(N_f,\Delta)=\fft{\pi\sqrt{2N_f\tDelta_1\tDelta_2\tDelta_3\tDelta_4}}{3}\,,
\end{equation}
read off from (\ref{ADHM:V:largeN}) with great precision for all configurations listed in (\ref{ADHM:configurations}). For a precise comparison, below we provide tables for the error ratio
\begin{equation}
	R_{3/2}(N_f,\Delta)\equiv\fft{\hat g_{3/2}^\text{(lmf)}(N_f,\Delta)-\hat g_{3/2}(N_f,\Delta)}{\hat g_{3/2}(N_f,\Delta)}
\end{equation}
and the numerical estimate for the constant term $\hat g_0^\text{(lmf)}(N_f,\Delta)$ together with the associated standard error $\sigma_0$.

\medskip

\noindent$\boldsymbol{\Delta=(\fft12,\fft12,1,0)}$
%
\begin{center}
	\footnotesize
	\begin{tabular}{ |c||c|c|c| } 
		\hline
		& $R_{3/2}$ & $\hat g_0^\text{(lmf)}$ & $\sigma_0$\\
		\hline\hline
		$N_f=1$ & $-2.198{\times}10^{-42}$ & $-0.18384102333840097854$ & $1.656{\times}10^{-39}$ \\
		\hline
		$N_f=2$ & $2.180{\times}10^{-31}$ & $-0.33723358961840868631$ & $2.233{\times}10^{-28}$\\
		\hline
		$N_f=3$ & $-4.117{\times}10^{-28}$ & $-0.62879449364922746645$ & $4.979{\times}10^{-25}$\\ 
		\hline
		$N_f=4$ & $-2.443{\times}10^{-23}$ & $-1.0432995069016606835$ & $3.268{\times}10^{-20}$\\ 
		\hline
	\end{tabular}
\end{center}
%

\noindent$\boldsymbol{\Delta=(\fft37,\fft47,1,0)}$
%
\begin{center}
	\footnotesize
	\begin{tabular}{ |c||c|c|c| } 
		\hline
		& $R_{3/2}$ & $\hat g_0^\text{(lmf)}$ & $\sigma_0$\\
		\hline\hline
		$N_f=1$ & $-5.100{\times}10^{-38}$ & $-0.18487381932915327786$ & $3.757{\times}10^{-35}$ \\
		\hline
		$N_f=2$ & $2.985{\times}10^{-28}$ & $-0.34306944663243885903$ & $2.943{\times}10^{-25}$\\
		\hline
		$N_f=3$ & $-2.948{\times}10^{-25}$ & $-0.64251698663465353444$ & $3.398{\times}10^{-22}$\\ 
		\hline
	\end{tabular}
\end{center}
%

\noindent$\boldsymbol{\Delta=(\fft38,\fft58,1,\fft{N_f}{10})}$
%
\begin{center}
	\footnotesize
	\begin{tabular}{ |c||c|c|c| } 
		\hline
		& $R_{3/2}$ & $\hat g_0^\text{(lmf)}$ & $\sigma_0$\\
		\hline\hline
		$N_f=1$ & $-9.517{\times}10^{-31}$ & $-0.18922588511715377025$ & $6.421{\times}10^{-28}$ \\
		\hline
		$N_f=2$ & $3.923{\times}10^{-23}$ & $-0.35418384892357442559$ & $3.414{\times}10^{-20}$\\
		\hline
		$N_f=3$ & $-1.193{\times}10^{-19}$ & $-0.666855367656914298975$ & $1.181{\times}10^{-16}$\\ 
		\hline
	\end{tabular}
\end{center}
%

\noindent$\boldsymbol{\Delta=(\fft1\pi,\fft2\pi,2-\fft3\pi,N_f\left(1-\fft3\pi\right))}$
%
\begin{center}
	\footnotesize
	\begin{tabular}{ |c||c|c|c| } 
		\hline
		& $R_{3/2}$ & $\hat g_0^\text{(lmf)}$ & $\sigma_0$\\
		\hline\hline
		$N_f=1$ & $-5.526{\times}10^{-31}$ & $-0.19164115564794187181$ & $3.703{\times}10^{-28}$ \\
		\hline
		$N_f=2$ & $3.275{\times}10^{-23}$ & $-0.39085989312606644267$ & $2.840{\times}10^{-20}$\\
		\hline
		$N_f=3$ & $-4.067{\times}10^{-20}$ & $-0.75585703584795923151$ & $4.020{\times}10^{-17}$\\ 
		\hline
		$N_f=4$ & $-9.539{\times}10^{-18}$ & $-1.2726405536526142820$ & $1.026{\times}10^{-14}$\\ 
		\hline
	\end{tabular}
\end{center}
For $\Delta=(\fft1\pi,\fft2\pi,2-\fft3\pi,N_f\left(1-\fft3\pi\right))$ with $N_f=3$, we confirmed explicitly that the choice of different $(\Delta_{\tq},\Delta_q)$ configurations in (\ref{ADHM:configurations}) does not affect the result.

\bibliographystyle{JHEP}
\bibliography{SCI}

\providecommand{\href}[2]{#2}\begingroup\raggedright\begin{thebibliography}{10}

\bibitem{Pestun:2007rz}
V.~Pestun, \emph{{Localization of gauge theory on a four-sphere and
  supersymmetric Wilson loops}},
  \href{http://dx.doi.org/10.1007/s00220-012-1485-0}{\emph{Commun. Math. Phys.}
  {\bf 313} (2012) 71--129}, [\href{https://arxiv.org/abs/0712.2824}{{\tt
  0712.2824}}].

\bibitem{Pestun:2016zxk}
V.~Pestun et~al., \emph{{Localization techniques in quantum field theories}},
  \href{http://dx.doi.org/10.1088/1751-8121/aa63c1}{\emph{J. Phys. A} {\bf 50}
  (2017) 440301}, [\href{https://arxiv.org/abs/1608.02952}{{\tt 1608.02952}}].

\bibitem{Kinney:2005ej}
J.~Kinney, J.~M. Maldacena, S.~Minwalla and S.~Raju, \emph{{An Index for 4
  dimensional super conformal theories}},
  \href{http://dx.doi.org/10.1007/s00220-007-0258-7}{\emph{Commun. Math. Phys.}
  {\bf 275} (2007) 209--254}, [\href{https://arxiv.org/abs/hep-th/0510251}{{\tt
  hep-th/0510251}}].

\bibitem{Romelsberger:2005eg}
C.~Romelsberger, \emph{{Counting chiral primaries in N = 1, d=4 superconformal
  field theories}},
  \href{http://dx.doi.org/10.1016/j.nuclphysb.2006.03.037}{\emph{Nucl. Phys. B}
  {\bf 747} (2006) 329--353}, [\href{https://arxiv.org/abs/hep-th/0510060}{{\tt
  hep-th/0510060}}].

\bibitem{Gutowski:2004ez}
J.~B. Gutowski and H.~S. Reall, \emph{{Supersymmetric AdS(5) black holes}},
  \href{http://dx.doi.org/10.1088/1126-6708/2004/02/006}{\emph{JHEP} {\bf 02}
  (2004) 006}, [\href{https://arxiv.org/abs/hep-th/0401042}{{\tt
  hep-th/0401042}}].

\bibitem{Gutowski:2004yv}
J.~B. Gutowski and H.~S. Reall, \emph{{General supersymmetric AdS(5) black
  holes}}, \href{http://dx.doi.org/10.1088/1126-6708/2004/04/048}{\emph{JHEP}
  {\bf 04} (2004) 048}, [\href{https://arxiv.org/abs/hep-th/0401129}{{\tt
  hep-th/0401129}}].

\bibitem{Chong:2005hr}
Z.~W. Chong, M.~Cvetic, H.~Lu and C.~N. Pope, \emph{{General non-extremal
  rotating black holes in minimal five-dimensional gauged supergravity}},
  \href{http://dx.doi.org/10.1103/PhysRevLett.95.161301}{\emph{Phys. Rev.
  Lett.} {\bf 95} (2005) 161301},
  [\href{https://arxiv.org/abs/hep-th/0506029}{{\tt hep-th/0506029}}].

\bibitem{Chong:2005da}
Z.~W. Chong, M.~Cvetic, H.~Lu and C.~N. Pope, \emph{{Five-dimensional gauged
  supergravity black holes with independent rotation parameters}},
  \href{http://dx.doi.org/10.1103/PhysRevD.72.041901}{\emph{Phys. Rev. D} {\bf
  72} (2005) 041901}, [\href{https://arxiv.org/abs/hep-th/0505112}{{\tt
  hep-th/0505112}}].

\bibitem{Kunduri:2006ek}
H.~K. Kunduri, J.~Lucietti and H.~S. Reall, \emph{{Supersymmetric multi-charge
  AdS(5) black holes}},
  \href{http://dx.doi.org/10.1088/1126-6708/2006/04/036}{\emph{JHEP} {\bf 04}
  (2006) 036}, [\href{https://arxiv.org/abs/hep-th/0601156}{{\tt
  hep-th/0601156}}].

\bibitem{Hosseini:2017mds}
S.~M. Hosseini, K.~Hristov and A.~Zaffaroni, \emph{{An extremization principle
  for the entropy of rotating BPS black holes in AdS$_{5}$}},
  \href{http://dx.doi.org/10.1007/JHEP07(2017)106}{\emph{JHEP} {\bf 07} (2017)
  106}, [\href{https://arxiv.org/abs/1705.05383}{{\tt 1705.05383}}].

\bibitem{Choi:2018hmj}
S.~Choi, J.~Kim, S.~Kim and J.~Nahmgoong, \emph{{Large AdS black holes from
  QFT}},  \href{https://arxiv.org/abs/1810.12067}{{\tt 1810.12067}}.

\bibitem{Cabo-Bizet:2018ehj}
A.~Cabo-Bizet, D.~Cassani, D.~Martelli and S.~Murthy, \emph{{Microscopic origin
  of the Bekenstein-Hawking entropy of supersymmetric AdS$_{5}$ black holes}},
  \href{http://dx.doi.org/10.1007/JHEP10(2019)062}{\emph{JHEP} {\bf 10} (2019)
  062}, [\href{https://arxiv.org/abs/1810.11442}{{\tt 1810.11442}}].

\bibitem{Benini:2018ywd}
F.~Benini and E.~Milan, \emph{{Black Holes in 4D $\mathcal{N}$=4
  Super-Yang-Mills Field Theory}},
  \href{http://dx.doi.org/10.1103/PhysRevX.10.021037}{\emph{Phys. Rev. X} {\bf
  10} (2020) 021037}, [\href{https://arxiv.org/abs/1812.09613}{{\tt
  1812.09613}}].

\bibitem{Cassani:2021fyv}
D.~Cassani and Z.~Komargodski, \emph{{EFT and the SUSY Index on the 2nd
  Sheet}}, \href{http://dx.doi.org/10.21468/SciPostPhys.11.1.004}{\emph{SciPost
  Phys.} {\bf 11} (2021) 004}, [\href{https://arxiv.org/abs/2104.01464}{{\tt
  2104.01464}}].

\bibitem{GonzalezLezcano:2020yeb}
A.~Gonz\'alez~Lezcano, J.~Hong, J.~T. Liu and L.~A. Pando~Zayas,
  \emph{{Sub-leading Structures in Superconformal Indices: Subdominant Saddles
  and Logarithmic Contributions}},
  \href{http://dx.doi.org/10.1007/JHEP01(2021)001}{\emph{JHEP} {\bf 01} (2021)
  001}, [\href{https://arxiv.org/abs/2007.12604}{{\tt 2007.12604}}].

\bibitem{ArabiArdehali:2021nsx}
A.~Arabi~Ardehali and S.~Murthy, \emph{{The 4d superconformal index near roots
  of unity and 3d Chern-Simons theory}},
  \href{http://dx.doi.org/10.1007/JHEP10(2021)207}{\emph{JHEP} {\bf 10} (2021)
  207}, [\href{https://arxiv.org/abs/2104.02051}{{\tt 2104.02051}}].

\bibitem{Honda:2019cio}
M.~Honda, \emph{{Quantum Black Hole Entropy from 4d Supersymmetric Cardy
  formula}}, \href{http://dx.doi.org/10.1103/PhysRevD.100.026008}{\emph{Phys.
  Rev. D} {\bf 100} (2019) 026008},
  [\href{https://arxiv.org/abs/1901.08091}{{\tt 1901.08091}}].

\bibitem{ArabiArdehali:2019tdm}
A.~Arabi~Ardehali, \emph{{Cardy-like asymptotics of the 4d $ \mathcal{N}=4 $
  index and AdS$_{5}$ blackholes}},
  \href{http://dx.doi.org/10.1007/JHEP06(2019)134}{\emph{JHEP} {\bf 06} (2019)
  134}, [\href{https://arxiv.org/abs/1902.06619}{{\tt 1902.06619}}].

\bibitem{Kim:2019yrz}
J.~Kim, S.~Kim and J.~Song, \emph{{A 4d $ \mathcal{N} $ = 1 Cardy Formula}},
  \href{http://dx.doi.org/10.1007/JHEP01(2021)025}{\emph{JHEP} {\bf 01} (2021)
  025}, [\href{https://arxiv.org/abs/1904.03455}{{\tt 1904.03455}}].

\bibitem{Cabo-Bizet:2019osg}
A.~Cabo-Bizet, D.~Cassani, D.~Martelli and S.~Murthy, \emph{{The asymptotic
  growth of states of the 4d $ \mathcal{N}=1 $ superconformal index}},
  \href{http://dx.doi.org/10.1007/JHEP08(2019)120}{\emph{JHEP} {\bf 08} (2019)
  120}, [\href{https://arxiv.org/abs/1904.05865}{{\tt 1904.05865}}].

\bibitem{Amariti:2019mgp}
A.~Amariti, I.~Garozzo and G.~Lo~Monaco, \emph{{Entropy function from toric
  geometry}},
  \href{http://dx.doi.org/10.1016/j.nuclphysb.2021.115571}{\emph{Nucl. Phys. B}
  {\bf 973} (2021) 115571}, [\href{https://arxiv.org/abs/1904.10009}{{\tt
  1904.10009}}].

\bibitem{ArabiArdehali:2019orz}
A.~Arabi~Ardehali, J.~Hong and J.~T. Liu, \emph{{Asymptotic growth of the 4d $
  \mathcal{N} $ = 4 index and partially deconfined phases}},
  \href{http://dx.doi.org/10.1007/JHEP07(2020)073}{\emph{JHEP} {\bf 07} (2020)
  073}, [\href{https://arxiv.org/abs/1912.04169}{{\tt 1912.04169}}].

\bibitem{Amariti:2020jyx}
A.~Amariti, M.~Fazzi and A.~Segati, \emph{{The SCI of $ \mathcal{N} $ = 4
  USp(2N$_{c}$) and SO(N$_{c}$) SYM as a matrix integral}},
  \href{http://dx.doi.org/10.1007/JHEP06(2021)132}{\emph{JHEP} {\bf 06} (2021)
  132}, [\href{https://arxiv.org/abs/2012.15208}{{\tt 2012.15208}}].

\bibitem{Ardehali:2021irq}
A.~A. Ardehali and J.~Hong, \emph{{Decomposition of BPS moduli spaces and
  asymptotics of supersymmetric partition functions}},
  \href{http://dx.doi.org/10.1007/JHEP01(2022)062}{\emph{JHEP} {\bf 01} (2022)
  062}, [\href{https://arxiv.org/abs/2110.01538}{{\tt 2110.01538}}].

\bibitem{GonzalezLezcano:2019nca}
A.~Gonz\'alez~Lezcano and L.~A. Pando~Zayas, \emph{{Microstate counting via
  Bethe Ans\"atze in the 4d $ \mathcal{N} $ = 1 superconformal index}},
  \href{http://dx.doi.org/10.1007/JHEP03(2020)088}{\emph{JHEP} {\bf 03} (2020)
  088}, [\href{https://arxiv.org/abs/1907.12841}{{\tt 1907.12841}}].

\bibitem{Lanir:2019abx}
A.~Lanir, A.~Nedelin and O.~Sela, \emph{{Black hole entropy function for toric
  theories via Bethe Ansatz}},
  \href{http://dx.doi.org/10.1007/JHEP04(2020)091}{\emph{JHEP} {\bf 04} (2020)
  091}, [\href{https://arxiv.org/abs/1908.01737}{{\tt 1908.01737}}].

\bibitem{Cabo-Bizet:2019eaf}
A.~Cabo-Bizet and S.~Murthy, \emph{{Supersymmetric phases of 4d $ \mathcal{N} $
  = 4 SYM at large $N$}},
  \href{http://dx.doi.org/10.1007/JHEP09(2020)184}{\emph{JHEP} {\bf 09} (2020)
  184}, [\href{https://arxiv.org/abs/1909.09597}{{\tt 1909.09597}}].

\bibitem{Cabo-Bizet:2020nkr}
A.~Cabo-Bizet, D.~Cassani, D.~Martelli and S.~Murthy, \emph{{The large-$N$
  limit of the 4d $ \mathcal{N} $ = 1 superconformal index}},
  \href{http://dx.doi.org/10.1007/JHEP11(2020)150}{\emph{JHEP} {\bf 11} (2020)
  150}, [\href{https://arxiv.org/abs/2005.10654}{{\tt 2005.10654}}].

\bibitem{Benini:2020gjh}
F.~Benini, E.~Colombo, S.~Soltani, A.~Zaffaroni and Z.~Zhang,
  \emph{{Superconformal indices at large $N$ and the entropy of AdS$_5$
  $\times$ SE$_5$ black holes}},
  \href{http://dx.doi.org/10.1088/1361-6382/abb39b}{\emph{Class. Quant. Grav.}
  {\bf 37} (2020) 215021}, [\href{https://arxiv.org/abs/2005.12308}{{\tt
  2005.12308}}].

\bibitem{Copetti:2020dil}
C.~Copetti, A.~Grassi, Z.~Komargodski and L.~Tizzano, \emph{{Delayed
  deconfinement and the Hawking-Page transition}},
  \href{http://dx.doi.org/10.1007/JHEP04(2022)132}{\emph{JHEP} {\bf 04} (2022)
  132}, [\href{https://arxiv.org/abs/2008.04950}{{\tt 2008.04950}}].

\bibitem{Goldstein:2020yvj}
K.~Goldstein, V.~Jejjala, Y.~Lei, S.~van Leuven and W.~Li, \emph{{Residues,
  modularity, and the Cardy limit of the 4d $ \mathcal{N} $ = 4 superconformal
  index}}, \href{http://dx.doi.org/10.1007/JHEP04(2021)216}{\emph{JHEP} {\bf
  04} (2021) 216}, [\href{https://arxiv.org/abs/2011.06605}{{\tt 2011.06605}}].

\bibitem{Cabo-Bizet:2020ewf}
A.~Cabo-Bizet, \emph{{From multi-gravitons to Black holes: The role of complex
  saddles}},  \href{https://arxiv.org/abs/2012.04815}{{\tt 2012.04815}}.

\bibitem{Choi:2021lbk}
S.~Choi, S.~Jeong and S.~Kim, \emph{{The Yang-Mills duals of small AdS black
  holes}},  \href{https://arxiv.org/abs/2103.01401}{{\tt 2103.01401}}.

\bibitem{Gaiotto:2021xce}
D.~Gaiotto and J.~H. Lee, \emph{{The Giant Graviton Expansion}},
  \href{https://arxiv.org/abs/2109.02545}{{\tt 2109.02545}}.

\bibitem{Choi:2021rxi}
S.~Choi, S.~Jeong, S.~Kim and E.~Lee, \emph{{Exact QFT duals of AdS black
  holes}},  \href{https://arxiv.org/abs/2111.10720}{{\tt 2111.10720}}.

\bibitem{Choi:2022ovw}
S.~Choi, S.~Kim, E.~Lee and J.~Lee, \emph{{From giant gravitons to black
  holes}},  \href{https://arxiv.org/abs/2207.05172}{{\tt 2207.05172}}.

\bibitem{Bobev:2022bjm}
N.~Bobev, V.~Dimitrov, V.~Reys and A.~Vekemans, \emph{{Higher-Derivative
  Corrections and AdS$_5$ Black Holes}},
  \href{https://arxiv.org/abs/2207.10671}{{\tt 2207.10671}}.

\bibitem{Cassani:2022lrk}
D.~Cassani, A.~Ruip\'erez and E.~Turetta, \emph{{Corrections to AdS$_5$ Black
  Hole Thermodynamics from Higher-Derivative Supergravity}},
  \href{https://arxiv.org/abs/2208.01007}{{\tt 2208.01007}}.

\bibitem{Bhattacharya:2008zy}
J.~Bhattacharya, S.~Bhattacharyya, S.~Minwalla and S.~Raju, \emph{{Indices for
  Superconformal Field Theories in 3,5 and 6 Dimensions}},
  \href{http://dx.doi.org/10.1088/1126-6708/2008/02/064}{\emph{JHEP} {\bf 02}
  (2008) 064}, [\href{https://arxiv.org/abs/0801.1435}{{\tt 0801.1435}}].

\bibitem{Bhattacharya:2008bja}
J.~Bhattacharya and S.~Minwalla, \emph{{Superconformal Indices for N = 6 Chern
  Simons Theories}},
  \href{http://dx.doi.org/10.1088/1126-6708/2009/01/014}{\emph{JHEP} {\bf 01}
  (2009) 014}, [\href{https://arxiv.org/abs/0806.3251}{{\tt 0806.3251}}].

\bibitem{Kim:2009wb}
S.~Kim, \emph{{The Complete superconformal index for N=6 Chern-Simons theory}},
  \href{http://dx.doi.org/10.1016/j.nuclphysb.2009.06.025}{\emph{Nucl. Phys. B}
  {\bf 821} (2009) 241--284}, [\href{https://arxiv.org/abs/0903.4172}{{\tt
  0903.4172}}].

\bibitem{Imamura:2011su}
Y.~Imamura and S.~Yokoyama, \emph{{Index for three dimensional superconformal
  field theories with general R-charge assignments}},
  \href{http://dx.doi.org/10.1007/JHEP04(2011)007}{\emph{JHEP} {\bf 04} (2011)
  007}, [\href{https://arxiv.org/abs/1101.0557}{{\tt 1101.0557}}].

\bibitem{Krattenthaler:2011da}
C.~Krattenthaler, V.~P. Spiridonov and G.~S. Vartanov, \emph{{Superconformal
  indices of three-dimensional theories related by mirror symmetry}},
  \href{http://dx.doi.org/10.1007/JHEP06(2011)008}{\emph{JHEP} {\bf 06} (2011)
  008}, [\href{https://arxiv.org/abs/1103.4075}{{\tt 1103.4075}}].

\bibitem{Kapustin:2011jm}
A.~Kapustin and B.~Willett, \emph{{Generalized Superconformal Index for Three
  Dimensional Field Theories}},  \href{https://arxiv.org/abs/1106.2484}{{\tt
  1106.2484}}.

\bibitem{Choi:2019zpz}
S.~Choi, C.~Hwang and S.~Kim, \emph{{Quantum vortices, M2-branes and black
  holes}},  \href{https://arxiv.org/abs/1908.02470}{{\tt 1908.02470}}.

\bibitem{Choi:2019dfu}
S.~Choi and C.~Hwang, \emph{{Universal 3d Cardy Block and Black Hole Entropy}},
  \href{http://dx.doi.org/10.1007/JHEP03(2020)068}{\emph{JHEP} {\bf 03} (2020)
  068}, [\href{https://arxiv.org/abs/1911.01448}{{\tt 1911.01448}}].

\bibitem{Nian:2019pxj}
J.~Nian and L.~A. Pando~Zayas, \emph{{Microscopic entropy of rotating
  electrically charged AdS$_{4}$ black holes from field theory localization}},
  \href{http://dx.doi.org/10.1007/JHEP03(2020)081}{\emph{JHEP} {\bf 03} (2020)
  081}, [\href{https://arxiv.org/abs/1909.07943}{{\tt 1909.07943}}].

\bibitem{GonzalezLezcano:2022hcf}
A.~Gonz\'alez~Lezcano, M.~Jerdee and L.~A. Pando~Zayas, \emph{{Cardy Expansion
  of 3d Superconformal Indices and Corrections to the Dual Black Hole
  Entropy}},  \href{https://arxiv.org/abs/2210.12065}{{\tt 2210.12065}}.

\bibitem{Aharony:2008ug}
O.~Aharony, O.~Bergman, D.~L. Jafferis and J.~Maldacena, \emph{{N=6
  superconformal Chern-Simons-matter theories, M2-branes and their gravity
  duals}}, \href{http://dx.doi.org/10.1088/1126-6708/2008/10/091}{\emph{JHEP}
  {\bf 10} (2008) 091}, [\href{https://arxiv.org/abs/0806.1218}{{\tt
  0806.1218}}].

\bibitem{Benini:2015noa}
F.~Benini and A.~Zaffaroni, \emph{{A topologically twisted index for
  three-dimensional supersymmetric theories}},
  \href{http://dx.doi.org/10.1007/JHEP07(2015)127}{\emph{JHEP} {\bf 07} (2015)
  127}, [\href{https://arxiv.org/abs/1504.03698}{{\tt 1504.03698}}].

\bibitem{Benini:2016hjo}
F.~Benini and A.~Zaffaroni, \emph{{Supersymmetric partition functions on
  Riemann surfaces}}, {\emph{Proc. Symp. Pure Math.} {\bf 96} (2017) 13--46},
  [\href{https://arxiv.org/abs/1605.06120}{{\tt 1605.06120}}].

\bibitem{Closset:2016arn}
C.~Closset and H.~Kim, \emph{{Comments on twisted indices in 3d supersymmetric
  gauge theories}},
  \href{http://dx.doi.org/10.1007/JHEP08(2016)059}{\emph{JHEP} {\bf 08} (2016)
  059}, [\href{https://arxiv.org/abs/1605.06531}{{\tt 1605.06531}}].

\bibitem{Bobev:2022jte}
N.~Bobev, J.~Hong and V.~Reys, \emph{{Large N Partition Functions, Holography,
  and Black Holes}},
  \href{http://dx.doi.org/10.1103/PhysRevLett.129.041602}{\emph{Phys. Rev.
  Lett.} {\bf 129} (2022) 041602},
  [\href{https://arxiv.org/abs/2203.14981}{{\tt 2203.14981}}].

\bibitem{Bobev:2022eus}
N.~Bobev, J.~Hong and V.~Reys, \emph{{Large $N$ Partition Functions of the ABJM
  Theory}},  \href{https://arxiv.org/abs/2210.09318}{{\tt 2210.09318}}.

\bibitem{Bobev:2023lkx}
N.~Bobev, J.~Hong and V.~Reys, \emph{{Large $N$ Partition Functions of 3d
  Holographic SCFTs}},  \href{https://arxiv.org/abs/2304.01734}{{\tt
  2304.01734}}.

\bibitem{Fuji:2011km}
H.~Fuji, S.~Hirano and S.~Moriyama, \emph{{Summing Up All Genus Free Energy of
  ABJM Matrix Model}},
  \href{http://dx.doi.org/10.1007/JHEP08(2011)001}{\emph{JHEP} {\bf 08} (2011)
  001}, [\href{https://arxiv.org/abs/1106.4631}{{\tt 1106.4631}}].

\bibitem{Marino:2011eh}
M.~Marino and P.~Putrov, \emph{{ABJM theory as a Fermi gas}},
  \href{http://dx.doi.org/10.1088/1742-5468/2012/03/P03001}{\emph{J. Stat.
  Mech.} {\bf 1203} (2012) P03001},
  [\href{https://arxiv.org/abs/1110.4066}{{\tt 1110.4066}}].

\bibitem{Mezei:2013gqa}
M.~Mezei and S.~S. Pufu, \emph{{Three-sphere free energy for classical gauge
  groups}}, \href{http://dx.doi.org/10.1007/JHEP02(2014)037}{\emph{JHEP} {\bf
  02} (2014) 037}, [\href{https://arxiv.org/abs/1312.0920}{{\tt 1312.0920}}].

\bibitem{Grassi:2014vwa}
A.~Grassi and M.~Marino, \emph{{M-theoretic matrix models}},
  \href{http://dx.doi.org/10.1007/JHEP02(2015)115}{\emph{JHEP} {\bf 02} (2015)
  115}, [\href{https://arxiv.org/abs/1403.4276}{{\tt 1403.4276}}].

\bibitem{Kostelecky:1995ei}
V.~A. Kostelecky and M.~J. Perry, \emph{{Solitonic black holes in gauged N=2
  supergravity}},
  \href{http://dx.doi.org/10.1016/0370-2693(95)01607-4}{\emph{Phys. Lett. B}
  {\bf 371} (1996) 191--198}, [\href{https://arxiv.org/abs/hep-th/9512222}{{\tt
  hep-th/9512222}}].

\bibitem{Caldarelli:1998hg}
M.~M. Caldarelli and D.~Klemm, \emph{{Supersymmetry of Anti-de Sitter black
  holes}}, \href{http://dx.doi.org/10.1016/S0550-3213(98)00846-3}{\emph{Nucl.
  Phys. B} {\bf 545} (1999) 434--460},
  [\href{https://arxiv.org/abs/hep-th/9808097}{{\tt hep-th/9808097}}].

\bibitem{Cvetic:2005zi}
M.~Cvetic, G.~W. Gibbons, H.~Lu and C.~N. Pope, \emph{{Rotating black holes in
  gauged supergravities: Thermodynamics, supersymmetric limits, topological
  solitons and time machines}},
  \href{https://arxiv.org/abs/hep-th/0504080}{{\tt hep-th/0504080}}.

\bibitem{Chow:2013gba}
D.~D.~K. Chow and G.~Comp\`ere, \emph{{Dyonic AdS black holes in maximal gauged
  supergravity}},
  \href{http://dx.doi.org/10.1103/PhysRevD.89.065003}{\emph{Phys. Rev. D} {\bf
  89} (2014) 065003}, [\href{https://arxiv.org/abs/1311.1204}{{\tt
  1311.1204}}].

\bibitem{Hristov:2019mqp}
K.~Hristov, S.~Katmadas and C.~Toldo, \emph{{Matter-coupled supersymmetric
  Kerr-Newman-AdS$_4$ black holes}},
  \href{http://dx.doi.org/10.1103/PhysRevD.100.066016}{\emph{Phys. Rev. D} {\bf
  100} (2019) 066016}, [\href{https://arxiv.org/abs/1907.05192}{{\tt
  1907.05192}}].

\bibitem{Choi:2018fdc}
S.~Choi, C.~Hwang, S.~Kim and J.~Nahmgoong, \emph{{Entropy Functions of BPS
  Black Holes in AdS$_{4}$ and AdS$_{6}$}},
  \href{http://dx.doi.org/10.3938/jkps.76.101}{\emph{J. Korean Phys. Soc.} {\bf
  76} (2020) 101--108}, [\href{https://arxiv.org/abs/1811.02158}{{\tt
  1811.02158}}].

\bibitem{Cassani:2019mms}
D.~Cassani and L.~Papini, \emph{{The BPS limit of rotating AdS black hole
  thermodynamics}},
  \href{http://dx.doi.org/10.1007/JHEP09(2019)079}{\emph{JHEP} {\bf 09} (2019)
  079}, [\href{https://arxiv.org/abs/1906.10148}{{\tt 1906.10148}}].

\bibitem{Bobev:2020egg}
N.~Bobev, A.~M. Charles, K.~Hristov and V.~Reys, \emph{{The Unreasonable
  Effectiveness of Higher-Derivative Supergravity in AdS$_4$ Holography}},
  \href{http://dx.doi.org/10.1103/PhysRevLett.125.131601}{\emph{Phys. Rev.
  Lett.} {\bf 125} (2020) 131601},
  [\href{https://arxiv.org/abs/2006.09390}{{\tt 2006.09390}}].

\bibitem{Bobev:2021oku}
N.~Bobev, A.~M. Charles, K.~Hristov and V.~Reys, \emph{{Higher-derivative
  supergravity, AdS$_{4}$ holography, and black holes}},
  \href{http://dx.doi.org/10.1007/JHEP08(2021)173}{\emph{JHEP} {\bf 08} (2021)
  173}, [\href{https://arxiv.org/abs/2106.04581}{{\tt 2106.04581}}].

\bibitem{Hristov:2021zai}
K.~Hristov and V.~Reys, \emph{{Factorization of log-corrections in
  AdS$_{4}$/CFT$_{3}$ from supergravity localization}},
  \href{http://dx.doi.org/10.1007/JHEP12(2021)031}{\emph{JHEP} {\bf 12} (2021)
  031}, [\href{https://arxiv.org/abs/2107.12398}{{\tt 2107.12398}}].

\bibitem{Hristov:2022lcw}
K.~Hristov, \emph{{ABJM at finite $N$ via 4d supergravity}},
  \href{https://arxiv.org/abs/2204.02992}{{\tt 2204.02992}}.

\bibitem{Aharony:2013dha}
O.~Aharony, S.~S. Razamat, N.~Seiberg and B.~Willett, \emph{{3d dualities from
  4d dualities}}, \href{http://dx.doi.org/10.1007/JHEP07(2013)149}{\emph{JHEP}
  {\bf 07} (2013) 149}, [\href{https://arxiv.org/abs/1305.3924}{{\tt
  1305.3924}}].

\bibitem{Dimofte:2011py}
T.~Dimofte, D.~Gaiotto and S.~Gukov, \emph{{3-Manifolds and 3d Indices}},
  \href{http://dx.doi.org/10.4310/ATMP.2013.v17.n5.a3}{\emph{Adv. Theor. Math.
  Phys.} {\bf 17} (2013) 975--1076},
  [\href{https://arxiv.org/abs/1112.5179}{{\tt 1112.5179}}].

\bibitem{Aharony:2013kma}
O.~Aharony, S.~S. Razamat, N.~Seiberg and B.~Willett, \emph{{3$d$ dualities
  from 4$d$ dualities for orthogonal groups}},
  \href{http://dx.doi.org/10.1007/JHEP08(2013)099}{\emph{JHEP} {\bf 08} (2013)
  099}, [\href{https://arxiv.org/abs/1307.0511}{{\tt 1307.0511}}].

\bibitem{Hristov:2021qsw}
K.~Hristov, \emph{{4d $ \mathcal{N} $ = 2 supergravity observables from
  Nekrasov-like partition functions}},
  \href{http://dx.doi.org/10.1007/JHEP02(2022)079}{\emph{JHEP} {\bf 02} (2022)
  079}, [\href{https://arxiv.org/abs/2111.06903}{{\tt 2111.06903}}].

\bibitem{Benini:2015eyy}
F.~Benini, K.~Hristov and A.~Zaffaroni, \emph{{Black hole microstates in
  AdS$_{4}$ from supersymmetric localization}},
  \href{http://dx.doi.org/10.1007/JHEP05(2016)054}{\emph{JHEP} {\bf 05} (2016)
  054}, [\href{https://arxiv.org/abs/1511.04085}{{\tt 1511.04085}}].

\bibitem{Hwang:2012jh}
C.~Hwang, H.-C. Kim and J.~Park, \emph{{Factorization of the 3d superconformal
  index}}, \href{http://dx.doi.org/10.1007/JHEP08(2014)018}{\emph{JHEP} {\bf
  08} (2014) 018}, [\href{https://arxiv.org/abs/1211.6023}{{\tt 1211.6023}}].

\bibitem{Hosseini:2016ume}
S.~M. Hosseini and N.~Mekareeya, \emph{{Large $N$ topologically twisted index:
  necklace quivers, dualities, and Sasaki-Einstein spaces}},
  \href{http://dx.doi.org/10.1007/JHEP08(2016)089}{\emph{JHEP} {\bf 08} (2016)
  089}, [\href{https://arxiv.org/abs/1604.03397}{{\tt 1604.03397}}].

\bibitem{Azzurli:2017kxo}
F.~Azzurli, N.~Bobev, P.~M. Crichigno, V.~S. Min and A.~Zaffaroni, \emph{{A
  universal counting of black hole microstates in AdS$_{4}$}},
  \href{http://dx.doi.org/10.1007/JHEP02(2018)054}{\emph{JHEP} {\bf 02} (2018)
  054}, [\href{https://arxiv.org/abs/1707.04257}{{\tt 1707.04257}}].

\bibitem{Bobev:2017uzs}
N.~Bobev and P.~M. Crichigno, \emph{{Universal RG Flows Across Dimensions and
  Holography}}, \href{http://dx.doi.org/10.1007/JHEP12(2017)065}{\emph{JHEP}
  {\bf 12} (2017) 065}, [\href{https://arxiv.org/abs/1708.05052}{{\tt
  1708.05052}}].

\bibitem{Bobev:2019zmz}
N.~Bobev and P.~M. Crichigno, \emph{{Universal spinning black holes and
  theories of class $ \mathcal{R} $}},
  \href{http://dx.doi.org/10.1007/JHEP12(2019)054}{\emph{JHEP} {\bf 12} (2019)
  054}, [\href{https://arxiv.org/abs/1909.05873}{{\tt 1909.05873}}].

\bibitem{deWit:1986oxb}
B.~de~Wit and H.~Nicolai, \emph{{The Consistency of the S**7 Truncation in D=11
  Supergravity}},
  \href{http://dx.doi.org/10.1016/0550-3213(87)90253-7}{\emph{Nucl. Phys. B}
  {\bf 281} (1987) 211--240}.

\bibitem{Gauntlett:2007ma}
J.~P. Gauntlett and O.~Varela, \emph{{Consistent Kaluza-Klein reductions for
  general supersymmetric AdS solutions}},
  \href{http://dx.doi.org/10.1103/PhysRevD.76.126007}{\emph{Phys. Rev. D} {\bf
  76} (2007) 126007}, [\href{https://arxiv.org/abs/0707.2315}{{\tt
  0707.2315}}].

\bibitem{Cvetic:1999xp}
M.~Cvetic, M.~J. Duff, P.~Hoxha, J.~T. Liu, H.~Lu, J.~X. Lu et~al.,
  \emph{{Embedding AdS black holes in ten-dimensions and eleven-dimensions}},
  \href{http://dx.doi.org/10.1016/S0550-3213(99)00419-8}{\emph{Nucl. Phys. B}
  {\bf 558} (1999) 96--126}, [\href{https://arxiv.org/abs/hep-th/9903214}{{\tt
  hep-th/9903214}}].

\bibitem{Azizi:2016noi}
A.~Azizi, H.~Godazgar, M.~Godazgar and C.~N. Pope, \emph{{Embedding of gauged
  STU supergravity in eleven dimensions}},
  \href{http://dx.doi.org/10.1103/PhysRevD.94.066003}{\emph{Phys. Rev. D} {\bf
  94} (2016) 066003}, [\href{https://arxiv.org/abs/1606.06954}{{\tt
  1606.06954}}].

\bibitem{Marino:2016new}
M.~Marino, \emph{{Localization at large N in Chern\textendash{}Simons-matter
  theories}}, \href{http://dx.doi.org/10.1088/1751-8121/aa5f69}{\emph{J. Phys.
  A} {\bf 50} (2017) 443007}, [\href{https://arxiv.org/abs/1608.02959}{{\tt
  1608.02959}}].

\bibitem{Closset:2019hyt}
C.~Closset and H.~Kim, \emph{{Three-dimensional N=2 supersymmetric gauge
  theories and partition functions on Seifert manifolds: A review}},
  \href{http://dx.doi.org/10.1142/S0217751X19300114}{\emph{Int. J. Mod. Phys.
  A} {\bf 34} (2019) 1930011}, [\href{https://arxiv.org/abs/1908.08875}{{\tt
  1908.08875}}].

\bibitem{Pasquetti:2019uop}
S.~Pasquetti and M.~Sacchi, \emph{{From 3$d$ dualities to 2$d$ free field
  correlators and back}},
  \href{http://dx.doi.org/10.1007/JHEP11(2019)081}{\emph{JHEP} {\bf 11} (2019)
  081}, [\href{https://arxiv.org/abs/1903.10817}{{\tt 1903.10817}}].

\end{thebibliography}\endgroup

\end{document}